\documentclass[a4paper, journal]{IEEEtran}
\usepackage{cite}
\usepackage{latexsym}
\usepackage{etoolbox}
\makeatletter
\patchcmd{\@makecaption}
  {\scshape}
  {}
  {}
  {}
\makeatletter
\patchcmd{\@makecaption}
  {\\}
  {.\ }
  {}
  {}
\makeatother
\def\tablename{Table}

\usepackage{tcolorbox}
\usepackage{color}
\usepackage{amsthm}
\usepackage{amsxtra}
\usepackage[mathscr]{euscript}
\usepackage{graphicx}
\graphicspath{{./original_source/}{./results_paper/plots/}{./results_smoke/plots/}}
\usepackage{multirow}
\usepackage{algpseudocode}
\usepackage{algorithmicx}
\usepackage{algpseudocode}
\usepackage{algpseudocode}
\usepackage{algorithm}
\usepackage{pifont}

\usepackage{rotating, graphicx}
\usepackage{array,multirow,graphicx}
\usepackage{amsmath}
\usepackage{tikz}
\usepackage{amssymb}
\usepackage{lipsum}
\usepackage{lipsum, multicol}
\usepackage{adjustbox}

\usepackage{upgreek}
\usepackage{mathtools}
\usepackage{float}
\usepackage{placeins}
\usepackage[table]{xcolor}
\definecolor{tableheader}{gray}{0.25}

\usepackage[caption = false]{subfig}
\usepackage{acronym}
\usepackage{orcidlink}
\usepackage{cuted} 
\usepackage{fancyhdr}
\hypersetup{hidelinks}
\usepackage{colortbl,booktabs}
\usepackage{svg}
\usepackage{supertabular}
\usepackage{comment}
\usepackage{alltt}
\usepackage{enumitem}
\usepackage{mathtools}
\usepackage{float}
\usepackage[caption = false]{subfig}
\usepackage{threeparttable}
\usepackage[utf8]{inputenc}
\usepackage[english]{babel}

\begin{document}

\IfFileExists{results_paper/result_macros.tex}{\input{results_paper/result_macros.tex}}{}
\providecommand{\NumberOfTestSeeds}{--}
\providecommand{\NumberOfTuningSeeds}{--}
\providecommand{\TabularSARSACompletion}{--}
\providecommand{\TabularSARSACost}{--}
\providecommand{\TabularSARSADecisionMs}{--}
\providecommand{\TabularQLearningCompletion}{--}
\providecommand{\TabularQLearningCost}{--}
\providecommand{\TabularQLearningDecisionMs}{--}
\providecommand{\LinearSARSACompletion}{--}
\providecommand{\LinearSARSACost}{--}
\providecommand{\LinearSARSADecisionMs}{--}
\providecommand{\LinearQLearningCompletion}{--}
\providecommand{\LinearQLearningCost}{--}
\providecommand{\LinearQLearningDecisionMs}{--}
\providecommand{\WastageAwareCompletion}{--}
\providecommand{\WastageAwareCost}{--}
\providecommand{\WastageAwareDecisionMs}{--}
\providecommand{\MILPCompletion}{--}
\providecommand{\MILPCost}{--}
\providecommand{\MILPDecisionMs}{--}

\begin{figure*}[ht] 
\centering
\begin{tcolorbox}[
colframe=black!75!black, 
colback=darkgray!5!white,   
coltitle=white,         
colbacktitle=darkgray!75!black, 
fonttitle=\bfseries, 
title= ,  
width=\textwidth,       
boxrule=0.3mm,            
sharp corners,          
left=1mm, right=1mm,    
top=1mm, bottom=1mm     
]
\textbf{Authors’ Note:} This is a preprint version of the article currently under review. We welcome constructive feedback and comments to further enhance the quality and clarity of this work. If you have any suggestions, insights, or critiques, please feel free to contact the corresponding author, Mohammad Asif Habibi\orcidlink{0000-0001-9874-0047}, at [\textcolor{blue}{mohammad\_asif.habibi@dfki.de}]. We sincerely appreciate all feedback, whether minor, critical, or substantial.
\\ \\ 

\textbf{Disclaimer:} This article has been submitted to the IEEE for possible publication. Copyright may be transferred without notice, after which this version may no longer be accessible. 
\\ \\ 

\textbf{Copyright:}
\textcopyright~$2026$~IEEE. Personal use of this material is permitted. If the article gets accepted, permission from IEEE must be obtained for all other uses in any current or future media, including reprinting/republishing this material for advertising or promotional purposes, creating new collective works, for resale or redistribution to servers or lists, or reuse of any copyrighted component of this work in other works.

\end{tcolorbox}
\vspace{-5.8mm} 
\end{figure*}

\title{\textbf{\texttt{SliceMapper}}: Intelligent Mapping of O-CU and O-DU onto O-Cloud Sites in 6G O-RAN}
\vspace{-0.40mm}

\author{Mohammad Asif Habibi\orcidlink{0000-0001-9874-0047}, 
Xavier Costa-P\'erez\orcidlink{0000-0002-9654-6109}~\IEEEmembership{(Senior Member,~IEEE)}, 
and Hans D. Schotten\orcidlink{0000-0001-5005-3635}~\IEEEmembership{(Member,~IEEE)}
\vspace{-11.8mm}
\thanks{\par\noindent\rule{0.99\columnwidth}{0.4pt}}
\thanks{This manuscript was received on DD MM $\mathrm{2027}$, revised on DD MM $\mathrm{2027}$, and accepted on DD MM $\mathrm{2027}$. It was published on DD MM $\mathrm{2027}$, and the current version was last updated on DD MM $\mathrm{2027}$. Mohammad Asif Habibi\orcidlink{0000-0001-9874-0047} and Hans D. Schotten\orcidlink{0000-0001-5005-3635} are with the Intelligent Networks (IN) Research Department, German Research Center for Artificial Intelligence (DFKI), $\mathrm{67663}$ Kaiserslautern, Germany. Xavier Costa-P\'erez\orcidlink{0000-0002-9654-6109} is with the Space and Terrestrial Systems Group, NEC Laboratories Europe, $\mathrm{69115}$ Heidelberg, Germany. He is also associated with the Department of AI-Driven Systems of the i$2$CAT Research Center and the Department of Engineering Sciences of the Catalan Institution for Research and Advanced Studies (ICREA), $\mathrm{08034}$ Barcelona, Spain. Hans D. Schotten\orcidlink{0000-0001-5005-3635} is also affiliated with the Division of Wireless Communications and Radio Navigation (WiCoN), Department of Electrical and Computer Engineering (EIT), University of Kaiserslautern (RPTU), $\mathrm{67663}$ Kaiserslautern, Germany. The corresponding author is Mohammad Asif Habibi\orcidlink{0000-0001-9874-0047} (\texttt{mohammad\_asif.habibi@dfki.de}).}
}
  
\maketitle
\begin{abstract}
In this paper, we propose an rApp, named \texttt{SliceMapper}, to optimize the mapping process of the open centralized unit (O-CU) and open distributed unit (O-DU) of an open radio access network (O-RAN) slice subnet onto the underlying open cloud (O-Cloud) sites in sixth-generation O-RAN. To that end, we first design a system model for \texttt{SliceMapper} and introduce its mathematical framework. Next, we formulate the mapping process addressed by \texttt{SliceMapper} as a capacity-constrained and feasibility-aware sequential decision-making optimization problem. This yields four learning variants: tabular state-action-reward-state-action (SARSA), tabular $Q$-learning, linear SARSA, and linear $Q$-learning. To evaluate the performance of these approaches, we conduct a series of simulations under various scenarios. Furthermore, we compare these learning variants against six baselines, namely random feasible placement, first-fit, first-fit decreasing, best-fit, a heuristic based on resource-wastage-awareness, and a time-bounded mixed-integer linear programming (MILP). The evaluation considers multiple mapping configurations over $30$ independent random seeds. The results demonstrate that in one of the mapping scenarios (namely the $16\times8$ scenario), the best reinforcement learning-based method achieves a placement fraction $\sim 0.996$, compared with the perfect $1.000$ for MILP. The results further exhibit that the linear function approximation significantly reduces representation-memory growth compared with the tabular methods, whereas the heuristics perform better in placement quality and decision latency. We also study the trade-offs among placement quality, decision latency, and representation cost. Finally, we discuss future research directions.
\end{abstract}

\IEEEoverridecommandlockouts
\begin{keywords}
5G, 6G, Non-RT RIC, O-Cloud Sites, O-CU, O-DU, O-RAN, O-RAN Slicing, Off-policy, On-policy SARSA, $Q$-Learning, rApps, Reinforcement Learning, SMO, \texttt{SliceMapper}
\vspace{-4.5mm}
\end{keywords}

\IEEEpeerreviewmaketitle

\section{Introduction}\label{Sec:Introduction}
\IEEEPARstart{T}{he} $mapping$ of \acp{VNF} onto the underlying physical infrastructure presents a complex $optimization$ problem \cite{10090468}. This problem has been extensively studied over the last decade in the context of cloud~computing \cite{IaaSPaper}. However, with the extension of $virtualization$ and $cloudification$ to the edge of cellular networks, combined with the unique characteristics of wireless communication $channels$ and the diverse $requirements$ of end-users and vertical industries, the \ac{VNF} mapping problem has become even more intricate \cite{9750106}. Furthermore, the recent introduction of $open$ interfaces and multi-vendor $interoperability$ in \ac{5G} and \ac{6G} \ac{O-RAN} has further amplified the complexity of \ac{VNF} mapping \cite{11124199}. Therefore, it is essential to reconsider traditional solutions and re-evaluate the properties of this optimization problem in light of these $new$ challenges.

In \ac{O-RAN}, the traditional $monolithic$ base station, namely \ac{O-gNB}, is decomposed into \ac{O-CU}, \ac{O-DU}, and \ac{O-RU}, which communicate over standardized open interfaces \cite{WG1ORANArchitecture}. The \ac{O-CU} and \ac{O-DU} (as \acp{VNF}) can be deployed on heterogeneous \ac{O-Cloud} sites and the \ac{O-RU} on cellular network sites. This architectural shift enables multi-vendor interoperability, flexible deployment, and $network~slicing$, but it also significantly increases the complexity of virtual and physical resource management \cite{9750106}. In particular, the mapping of \ac{O-CU} and \ac{O-DU} onto available \acp{VM} in \ac{O-Cloud} sites must account for diverse computational and storage requirements, stringent performance constraints, and dynamic network conditions. Hence, traditional static or rule-based mapping $strategies$ might face limitations in efficiently handling the scale and variability of \ac{O-RAN} deployments beyond \ac{5G} and \ac{6G} \cite{9750106}. 

The challenge becomes even more intensified in \ac{O-RAN} slicing, where multiple slice subnets with diverse service requirements coexist on a shared wireless infrastructure \cite{9750106}. Each subnet may impose different latency, throughput, and reliability constraints, requiring intelligent and adaptive mapping of \acp{VNF} to avoid resource under-utilization or congestion \cite{9750106}. Moreover, the \ac{O-RAN} \ac{SMO} framework introduces the concept of rApps operating in the \ac{Non-RT RIC} \cite{WG2ORANNearRTRIC}, enabling data-driven and policy-based optimization over long time scales. While this creates an opportunity to leverage \ac{AI}/\ac{ML} for \ac{O-RAN} optimization, it also raises fundamental questions regarding how learning-based approaches should be designed, trained, deployed, and evaluated for complex \ac{VNF} mappings in arbitrarily large discrete action representations\cite{10604823}.

\subsection{Prior Works}\label{subsub:priorworks}
To this end, several methodologies have been proposed to study \ac{VNF} mapping in both cloud/data centers and wireless networks, including \ac{AI}/\ac{ML}-based approaches. A comprehensive comparison of representative state-of-the-art works is provided in Appendix~\ref{App:StateOfArt}. Notably, existing studies generally model each \ac{VNF} as a monolithic entity and investigate its mapping onto the available underlying \acp{PM}. Furthermore, most works primarily address \ac{VM}-to-\ac{PM} placement problems. Hence, they typically virtualize the compute and storage resources of \acp{PM} first and subsequently allocate these virtualized resources to \acp{VNF} in the form of \acp{VM} or containers. It is worth noting that the contribution here is not a new temporal-difference method. It is a \ac{VNFC}-to-\ac{VM} level formulation defined over compute and storage resource constraints.

\subsection{Major Research Challenges} \label{Subsec:Problem}
However, \textbf{the monolithic treatment of \acp{VNF} during the mapping process}, particularly in \ac{O-RAN} slicing, introduces several additional limitations. First, each \ac{VNF} within an \ac{O-RAN} slice \textbf{typically comprises multiple micro-functionalities} \cite{9750106}, each of which may require different amounts of virtual compute and storage resources to execute a given set of tasks. Second, \textbf{the execution requirements of these micro-functionalities may vary significantly} across different subnets \cite{8931318}. For example, a specific functionality may be disabled or require substantially more or fewer resources in one type of subnet compared to another. Consequently, modeling and mapping a \ac{VNF} as a monolithic entity, specifically across various types of \ac{O-RAN} slices, \textbf{becomes increasingly inefficient and difficult to scale}. Moreover, decomposing the mapping problem to the \ac{VNFC} level introduces a problem of not only obtaining feasible placements, but also efficiently and scalably doing so. Therefore, this optimization problem warrants re-examination within \ac{O-RAN} slicing.

\subsection{Design Goals and System Objectives}\label{Subsec:GoalsandObjectives}
To address the aforementioned challenges, we adopt a strategy that departs from existing methodologies and is grounded in the following design choices. First, \textbf{we decompose each \ac{VNF} (i.e., \ac{O-CU} and \ac{O-DU}) of an \ac{O-RAN} slice into multiple micro-functionalities}. These micro-functionalities are referred to as \acp{VNFC}, following the terminology defined by the \ac{ETSI} \cite{9750106}. While the \ac{VNFC}-to-\ac{VM} decomposition of the placement problem introduces additional design considerations, it also enables finer-grained control over resource allocation; its rationale and implications are discussed in Appendix~\ref{App:VNFCJustification}. 

Second, we consider \textbf{the virtual resource requirements of each \ac{VNFC}} and select \textbf{a \ac{VM} from a candidate set} onto which the considered \ac{VNFC} is mapped, which makes it a \ac{VNFC}-to-\ac{VM} placement problem.  Third, \textbf{we assume that the \ac{VNFC}-to-\ac{VM} mapping decisions are made by an rApp deployed in the \ac{Non-RT RIC} and executed by the resource manager}. This rApp is referred to as \texttt{SliceMapper}. 

Finally, \textbf{we formulate the \ac{VNFC}-to-\ac{VM} mapping task as a sequential decision-making optimization problem}, as justified in Appendix~\ref{App:IsMappingaSDMP?}. To address this challenge, we investigate both on-policy \ac{SARSA} and off-policy $Q$-learning, each in both linear and tabular forms, which are further compared against the conventional placement heuristics and an \ac{MILP} baseline. Scalability, empirical stabilization, computational cost, and failure conditions are examined separately. The goal of the \texttt{SliceMapper} is to ensure feasible, efficient, and resource-aware placements across \acp{VM} within \acp{O-Cloud}.

\subsection{Summary of Key Contributions}\label{Subsec:Contributions}
In line with the objectives outlined above, the main contributions of this paper are summarized as follows:

\begin{itemize}

    \item We \textbf{design an \ac{O-RAN}-oriented system model for} \texttt{SliceMapper}, which is an rApp-based framework that addresses the mapping of \acp{VNFC} of the \ac{O-CU} and \ac{O-DU} within an \ac{O-RAN} slice onto \acp{VM} within \acp{O-Cloud}.

    \item We \textbf{analyze four \ac{RL} methods}, covering both tabular and linear \ac{SARSA}, and tabular and linear $Q$-learning, within a capacity-constrained, feasibility-aware sequential placement environment.

    \item We \textbf{perform an extensive simulation-based comparative study of the four learning approaches} with random feasible placement, first-fit, first-fit decreasing, best fit, a resource-wastage-aware heuristic, and an \ac{MILP} baseline. We independently tuned the hyperparameters, multiple controlled problem sizes, placement quality, efficiency, decision runtime, training runtime, memory consumption, and tabular representation growth.

    \item We \textbf{analyze the strengths and weaknesses of the \ac{RL} methods} by performing stability and scalability analysis. The results show that \ac{RL} can achieve high placed-\ac{VNFC} fractions while being latency-efficient, whereas strong conventional heuristics remain highly competitive, and the \ac{MILP} baseline achieves the best static efficiency at substantially higher computational cost. The analysis further demonstrates the memory-scaling advantage of the linear representation methods over the tabular methods.

\end{itemize}

\subsection{The Structure of the Paper}\label{Subsec:Structure}
The remainder of this article is organized as follows: Section~\ref{Sec:SystemModel} presents the system model for \texttt{SliceMapper}. Section~\ref{Sec:ProblemFormulation} thoroughly formulates and studies the capacity-constrained \ac{VNFC}-to-\ac{VM} optimization problem. Section~\ref{Sec:ProposedSolution} defines the sequential learning formulation of the optimization problem in the context of \ac{RL} and the four considered \ac{RL} methods. Section~\ref{Sec:PerformanceEvaluationRevisedFinal} conducts an empirical stability analysis, comparison with the aforementioned baselines, scalability analysis, and reward ablation. Finally, Section~\ref{Sec:Conclusion} concludes the paper and outlines directions for future work.

\begin{figure*}[ht]
    \centering
    \includegraphics[height=3.3 in, width=7.15 in]{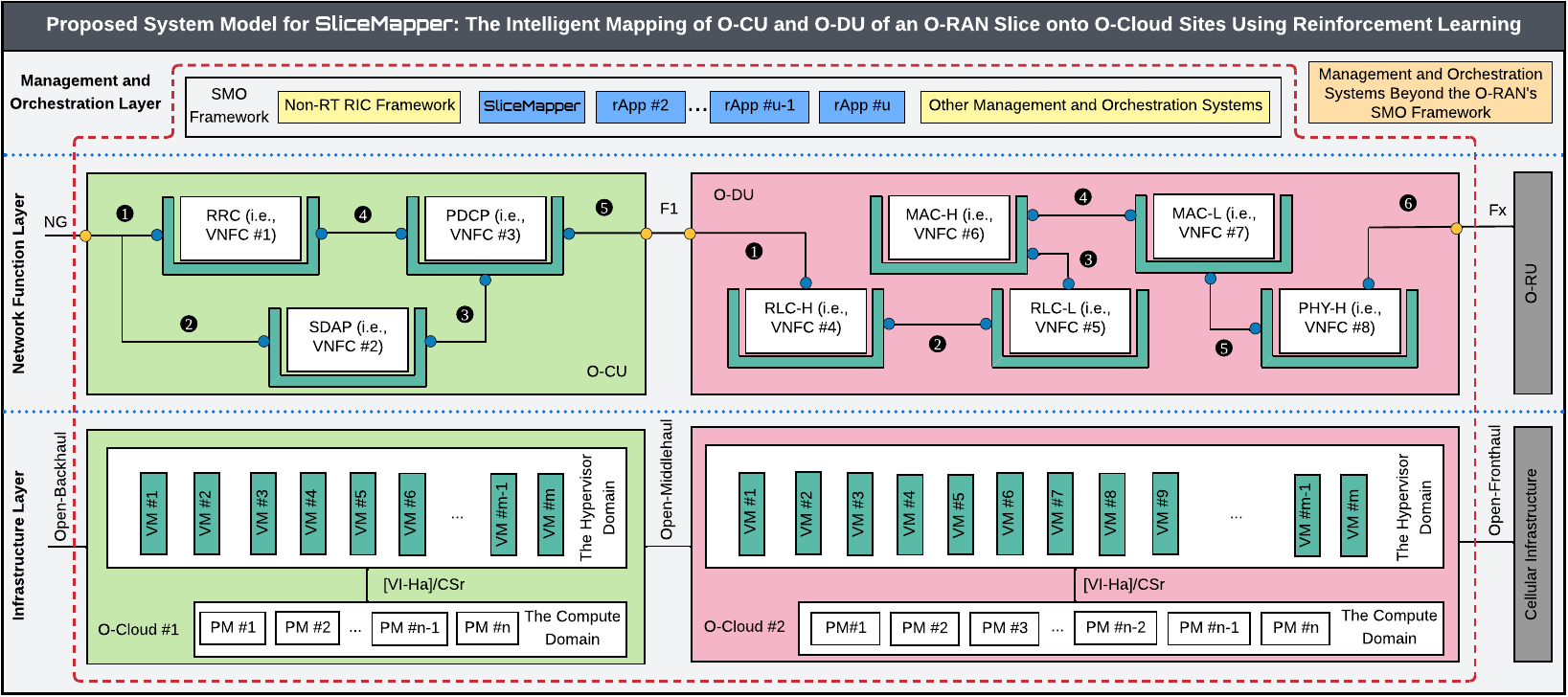}
    \vspace{-5.5mm}
    \caption{Proposed system model for mapping the \acp{VNFC} of the \ac{O-CU} and \ac{O-DU} of an \ac{O-RAN} slice onto \acp{VM} within \acp{O-Cloud}. Note that only the \acp{VNFC} of an \ac{O-gNB} and their corresponding virtual resources within \acp{O-Cloud}, illustrated in the red-dashed box, are within the scope of this article.}
    \label{Fig:SystemModel}
    \vspace{-4.0mm}
\end{figure*}


\section{System Model for \texttt{SliceMapper}}\label{Sec:SystemModel}
Figure $\mathrm{\ref{Fig:SystemModel}}$ illustrates the composition of the proposed system model for \texttt{SliceMapper} into three distinct $layers$: the Management and Orchestration Layer, the Network Function Layer, and the Infrastructure Layer. The Management and Orchestration Layer includes the \ac{SMO} framework, which manages and orchestrates \ac{O-RAN} components within a control loop operating on a timescale $t$ $\geq1$ $\mathrm{second(s)}$ \cite{9750106}. This layer primarily incorporates the \ac{Non-RT RIC} Framework, \texttt{SliceMapper}, and other authorized rApps. The Network Function Layer delivers the functionalities of an \ac{O-gNB}, comprising an \ac{O-CU}, \ac{O-DU}, and an \ac{O-RU}, along with the internal and external logical links of the \ac{O-gNB}. The Infrastructure Layer consists of \ac{O-Cloud} \#$\mathrm{1}$, \ac{O-Cloud} \#$\mathrm{2}$, and the cellular network infrastructure. These sites host the \ac{O-CU}, \ac{O-DU}, and \ac{O-RU}, respectively. Each \ac{O-Cloud} site consists of a set of \acp{VM} that are $abstracted$ from the underlying \acp{PM} by the Hypervisor Domain. Each \ac{VM} has a corresponding virtual compute and storage capacity and may host $\geq 1$ \acp{VNFC} of the \ac{O-CU} and \ac{O-DU}, if the corresponding residual resource capacities are sufficient. The optimization problem solves the mapping problem at \ac{VNFC}-to-\ac{VM} level. In the following subsections, we present mathematical descriptions for the three layers of the proposed system model. Key notations related to the system model are summarized in Table $\mathrm{\ref{Tab:SMNotations}}$.

\subsection{Modeling the Network Function Layer}\label{Subsec:NFLModeling}
Let $\mathcal{F}$ = \{$\mathit{f}$\textsubscript{1}, $\mathit{f}$\textsubscript{2}, ..., $\mathit{f}$\textsubscript{\textit{h}-1}, $\mathit{f}$\textsubscript{\textit{h}}\} be a set of $\mathit{h}$ physical and virtual functions of an \ac{O-RAN} slice subnet. These functions are processed sequentially by the \ac{O-CU}, \ac{O-DU}, and \ac{O-RU} in both upstream and downstream directions \cite{9750106}. Let $\mathcal{C}$ = \{$\mathit{f}$\textsubscript{1}, $\mathit{f}$\textsubscript{2}, $\mathit{f}$\textsubscript{3}\} ${\displaystyle \subset}$ $\mathcal{F}$ be a subset of \acp{VNFC} assigned to the \ac{O-CU}. In turn, let $\mathcal{D}$ = \{$\mathit{f}$\textsubscript{4}, $\mathit{f}$\textsubscript{5}, $\mathit{f}$\textsubscript{6}, $\mathit{f}$\textsubscript{7}, $\mathit{f}$\textsubscript{8}\} ${\displaystyle \subset}$ $\mathcal{F}$ be a subset of \acp{VNFC} assigned to the \ac{O-DU}. The subset $\mathcal{R} \subset \mathcal{F}$ constitutes the remaining functions of $\mathcal{F}$. These functions are the physical functions processed in the \ac{O-RU}. Consequently, we have $\mathcal{C}\cup\mathcal{D}\cup\mathcal{R}=\mathcal{F}$. It is worth noting that $\mathcal{C}$, $\mathcal{D}$, and $\mathcal{R}$ are pairwise disjoint sets, meaning their intersection is an empty set ($\varnothing$). Additionally, the details of $\mathcal{R}$ are beyond the scope. Therefore, hereafter by $\mathcal{F}$, we always mean $\mathcal{F} = \mathcal{C} \cup \mathcal{D}$, and the index $h$ does not include the physical \ac{O-RU} functions.

Functions \textit{f}\textsubscript{1}, \textit{f}\textsubscript{2}, and \textit{f}\textsubscript{3} correspond to the \ac{RRC}, \ac{SDAP}, and \ac{PDCP}, respectively. Functions \textit{f}\textsubscript{4}, \textit{f}\textsubscript{5}, \textit{f}\textsubscript{6}, \textit{f}\textsubscript{7}, and \textit{f}\textsubscript{8} represent the \ac{RLC}-High, \ac{RLC}-Low, \ac{MAC}-High, \ac{MAC}-Low, and \ac{PHY}-High, respectively. Each \ac{VNFC} is assumed to be hosted by a $single$ \ac{VM} (or by a Unikernel or a Container). However, a candidate \ac{VM} may host multiple \acp{VNFC}, provided that sufficient residual compute and storage capacities remain after each placement. 

Each \ac{VNFC}, $f_i\in \mathcal{C}\cup\mathcal{D}$ requires a certain amount of virtual compute and storage resources, denoted by $C_{i}^{\rm req}$ and $S_{i}^{\rm req}$, respectively. Let $\mathcal{V}=\{v_1,v_2,\ldots,v_m\}$ be the set of $m$ \acp{VM}. Denote the residual compute and storage capacities of \ac{VM} $v_j$ by $r_{j,t}^{C}$ and $r_{j,t}^{S}$, where $t$ denotes the placement step. A \ac{VNFC} $f_i$ can only be placed onto $v_j$ at placement step $t$ if and only if \begin{equation} C_i^{\rm req}\leq r_{j,t}^{C} \end{equation} and \begin{equation} S_i^{\rm req}\leq r_{j,t}^{S}. \end{equation} Upon successful completion of the placement, the residual capacities of the selected \ac{VM} $v_j$ are updated as follows \begin{equation} r_{j,t+1}^{C}=r_{j,t}^{C}-C_i^{\rm req} \end{equation} and \begin{equation} r_{j,t+1}^{S}=r_{j,t}^{S}-S_i^{\rm req}. \end{equation}

The total required virtual compute and storage resources of an \ac{O-CU} and \ac{O-DU} are obtained by: \begin{align} \mathbb{T}_{\rm com}^{\rm req}  =   \underbrace{\sum\limits_{{f_i \in \mathcal{C}}} {C_i^{\rm req}}}_\text{compute resource of O-CU} + \overbrace{\sum\limits_{{f_i \in \mathcal{D}}} {C_i^{\rm req}}}^{\text{compute resource of O-DU}} \end{align} and \begin{align}
  \mathbb{T}_{\rm sto}^{\rm req} = 
    \underbrace{\sum\limits_{{f_i \in \mathcal{C}}} {S_i^{\rm req}}}_\text{storage resource of O-CU} +
    \overbrace{\sum\limits_{{f_i \in \mathcal{D}}} {S_i^{\rm req}}}^{\text{storage resource of O-DU}},
\end{align}
respectively. Throughout this and the following discussion, the storage and compute capacities and requirements are always expressed in synthetic rather than physical resource units.

\subsection{Modeling the Infrastructure Layer}\label{Subsec:ILModeling}
Let $\mathcal{V}$ = \{$\mathit{v}$\textsubscript{1}, $\mathit{v}$\textsubscript{2}, ..., $\mathit{v}$\textsubscript{\textit{m}-1}, $\mathit{v}$\textsubscript{\textit{m}}\} be a set of $\mathit{m}$ \acp{VM} in the Hypervisor Domain, and let $\mathcal{P}$ = \{$\mathit{p}$\textsubscript{1}, $\mathit{p}$\textsubscript{2}, ..., $\mathit{p}$\textsubscript{\textit{n}-1}, $\mathit{p}$\textsubscript{\textit{n}}\} be a set of $\mathit{n}$ \acp{PM} in the Compute Domain. The set $\mathcal{V}$ is abstracted from set $\mathcal{P}$ in both \ac{O-Cloud} \#$\mathrm{1}$ and \ac{O-Cloud} \#$\mathrm{2}$. Each \ac{O-Cloud} site can consist of a large number of \acp{PM}, with each \ac{PM} capable of $hosting$ a specific number of \acp{VM}. 

Each \ac{VM}, $v_j\in$ $\mathcal{V}$ is characterized by a vector of available virtual resources. This means that each $v_j\in$ $\mathcal{V}$ has a maximum virtual compute capacity, denoted by $C_{j}^{\rm cap}$, and a maximum virtual storage capacity, denoted by $S_{j}^{\rm cap}$. The residual capacities at a time step $t$ are $r_{j,t}^{C}$ and $r_{j,t}^{S}$, which combinedly are essential in deciding whether the current \ac{VNFC} can be feasibly assigned to $v_j$. Note that the architecture model is allowed to permit a single \ac{VM} to host multiple \acp{VNFC}, subject to the capacity constraints formally introduced in Section~\ref{Sec:ProblemFormulation}.

\begin{table}[t]
\caption{List of key notations related to the proposed system model}
\centering
\renewcommand{\arraystretch}{1.4}
\begin{tabular}{|l|m{6.9cm}|}
\hline 
\cline{1-2}
\multicolumn{1}{|c|}{\cellcolor[HTML]{343434}\textbf{\textcolor{white}{Notation}}} & \multicolumn{1}{c|}{\cellcolor[HTML]{343434} \textbf{\textcolor{white}{Description}}}  \\ \cline{1-2}
\hline \hline 

$\mathcal{F}$ & Set of $h$ \ac{O-CU} and \ac{O-DU} \acp{VNFC}, i,e., $\mathcal{C}\cup \mathcal{D}$  \\ \cline{1-2}

$\mathcal{C}$  & Set of \acp{VNFC} assigned to \ac{O-CU} \\ \cline{1-2}

$\mathcal{D}$ & Set of \acp{VNFC} assigned to \ac{O-DU} \\ \cline{1-2}

$\mathcal{R}$ & Set of physical functions assigned to \ac{O-RU} \\ \cline{1-2}

$\mathcal{V}$ & Set of $m$ \acp{VM} in the Hypervisor Domain \\ \cline{1-2}

$\mathcal{P}$ & Set of $n$ \acp{PM} in the Compute Domain \\ \cline{1-2}

$\mathcal{Z}$ & Set of \textit{u} rApps in the \ac{Non-RT RIC} \\ \cline{1-2}

$\mathcal{O}$ & Set of \textit{d} tasks that are executed by a set of \textit{u} rApps in $\mathcal{Z}$ \\ \cline{1-2}

$C_i^{\rm req}$ & Virtual compute resources required by \ac{VNFC} $f_i\in \mathcal{F}$ \\ \cline{1-2}

$S_i^{\rm req}$ & Virtual storage resources required by \ac{VNFC} $f_i\in \mathcal{F}$ \\ \cline{1-2}

$\mathbb{T}_{\rm com}^{\rm req}$ & Total required virtual compute resources of an \ac{O-RAN} slice \\ \cline{1-2}

$\mathbb{T}_{\rm sto}^{\rm req}$ & Total required virtual storage resources of an \ac{O-RAN} slice \\ \cline{1-2}

$C_{k}^{\rm max}$ & Maximum physical compute capacity of \ac{PM} $p_k\in$ $\mathcal{P}$ \\ \cline{1-2}

$S_{k}^{\rm max}$ & Maximum physical storage capacity of \ac{PM} $p_k\in$ $\mathcal{P}$ \\ \cline{1-2}

$C_{k}^{\rm ava}$ & Available compute resources in \ac{PM} $p_k\in$ $\mathcal{P}$ \\ \cline{1-2}

$S_{k}^{\rm ava}$ & Available storage resources in \ac{PM} $p_k\in$ $\mathcal{P}$ \\ \cline{1-2}

$C_{j}^{\rm cap}$ & Available virtual compute capacity in \ac{VM} $v_j\in$ $\mathcal{V}$ \\ \cline{1-2}

$S_{j}^{\rm cap}$ & Available virtual storage capacity in \ac{VM} $v_j\in$ $\mathcal{V}$ \\ \cline{1-2}

$W$ & normalized residual resource function \\ \cline{1-2}

\end{tabular}
\label{Tab:SMNotations}
\vspace{2.5mm}
\end{table}

Likewise, each \ac{PM}, $p_k\in$ $\mathcal{P}$ has a maximum physical compute capacity, denoted by $C_{k}^{\rm max}$, and a maximum physical storage capacity, denoted by $S_{k}^{\rm max}$. The former represents the physical resource capacity for computing, and the latter represents the physical resource capacity for storing the \ac{VM}, $v_j\in$ $\mathcal{V}$ at a given time. These quantities define the architectural context for virtualizing candidate \acp{VM} but do not directly enter the \ac{RL} context of \texttt{SliceMapper}.

\subsection{Modeling the Management and Orchestration Layer}\label{Subsec:MOLModeling}
In this layer, alongside other management and orchestration components that may exist within or beyond the \ac{SMO} framework, a large set of rApps can also be placed. Let $\mathcal{Z}$ = \{$\mathit{z}$\textsubscript{1}, $\mathit{z}$\textsubscript{2}, ..., $\mathit{z}$\textsubscript{\textit{u}-1}, $\mathit{z}$\textsubscript{\textit{u}}\} represent a set of \textit{u} rApps within the \ac{SMO} framework. Each rApp $z_l\in$ $\mathcal{Z}$ enables the intelligent control and optimization of certain tasks, providing policy-based guidance to the components of the \ac{O-RAN} architecture on a control loop with a timescale of $t\geq1$ $\mathrm{second}$. We assume there exists a set $\mathcal{O}$ = \{$\mathit{o}$\textsubscript{1}, $\mathit{o}$\textsubscript{2}, ..., $\mathit{o}$\textsubscript{\textit{d}-1}, $\mathit{o}$\textsubscript{\textit{d}}\} of the intelligent control and optimization tasks within \ac{O-RAN} that can be performed by set $\mathcal{Z}$. Each optimization task $o_g\in$ $\mathcal{O}$ is executed by at least one rApp $z_l\in$ $\mathcal{Z}$.

From a task execution perspective, we believe there could be two types of rApps in set $\mathcal{Z}$: Single-Task and Multi-Task.
\begin{itemize}
    \item Single-Task rApps perform one specific optimization task of $o_g\in$ $\mathcal{O}$. They focus on delivering a specialized, streamlined experience, often excelling in their particular domain. These rApps provide simple and efficient tools for users requiring a specific functionality, without the complexity of additional features.
    \item Multi-Task rApps execute multiple optimization tasks, included in set $\mathcal{O}$, within the \ac{Non-RT RIC}. They integrate various features, allowing users to accomplish a range of activities without needing to switch between different rApps. These rApps are often versatile, catering to diverse user needs, and they aim to enhance productivity by consolidating multiple tasks into a unified user experience.
\end{itemize}

\texttt{SliceMapper}, which is placed in the Management and Orchestration Layer, as illustrated in Figure $\mathrm{\ref{Fig:SystemModel}}$, is a member of set $\mathcal{Z}$. It can function as either a Single-Task or a Multi-Task rApp, performing a certain number of tasks from set $\mathcal{O}$. Its goal is to find an efficient and feasible solution for mapping the \acp{VNFC} in subsets $\mathcal{C}$ and $\mathcal{D}$ of set $\mathcal{F}$ onto specific \acp{VM} in set $\mathcal{V}$, such that a predefined set of optimization objectives and a set of constraints (discussed in Section $\mathrm{\ref{Sec:ProblemFormulation}}$) are fulfilled. \texttt{SliceMapper} is not responsible for placing the \acp{VM} in set $\mathcal{V}$ onto \acp{PM} in set $\mathcal{P}$. We consider this task, which has been extensively examined in the literature \cite{IaaSPaper}, to be executed by a resource manager, such as the \ac{VIM} in \ac{NFV-MANO}. However, the resource manager can leverage the output of \texttt{SliceMapper} to efficiently and intelligently partition the resources of \acp{PM} in set $\mathcal{P}$, dynamically allocating only the required virtual resources to \acp{VM} in set $\mathcal{V}$ to host the corresponding \acp{VNFC} of an \ac{O-RAN} slice in subsets $\mathcal{C}$ and $\mathcal{D}$ of set $\mathcal{F}$. As a result, the output of \texttt{SliceMapper} helps prevent both over-utilization and under-utilization of resources in \acp{O-Cloud}.


\begin{table}
\caption{List of key notations related to the proposed RL-assisted framework}
\centering
\renewcommand{\arraystretch}{1.4}
\begin{tabular}{|l|m{6.9cm}|}
\hline
\cline{1-2}
\multicolumn{1}{|c|}{\cellcolor[HTML]{343434}\textbf{\textcolor{white}{Notation}}}
&
\multicolumn{1}{c|}{\cellcolor[HTML]{343434}\textbf{\textcolor{white}{Description}}}
\\ \cline{1-2}
\hline \hline

$\mathcal{S}$ &
State space, the set of all possible states in which \texttt{SliceMapper} can exist within the environment
\\ \cline{1-2}

$S_t$ &
State at decision step $t$
\\ \cline{1-2}

$s$ &
Arbitrary state in the state space $\mathcal{S}$
\\ \cline{1-2}

$\mathcal{A}(s)$ &
Set of candidate actions available at state $s$
\\ \cline{1-2}

$\mathcal{A}_{f}(s)$ &
Set of all feasible \acp{VM} that can be selected at state $s$
\\ \cline{1-2}

$A_t$ &
Action selected at decision step $t$
\\ \cline{1-2}

$a$ &
Arbitrary action in the action space
\\ \cline{1-2}

$R_t$ &
Observed reward after decision at step $t-1$
\\ \cline{1-2}

$G_t$ &
Discounted return starting at the decision step $t$
\\ \cline{1-2}

$\pi(a|s)$ &
Policy, i.e., the probability of selecting action $a$ in state $s$
\\ \cline{1-2}

$\pi^*$ &
Optimal policy maximizing expected return
\\ \cline{1-2}

$Q_\pi(s,a)$ &
State-action value under policy $\pi$
\\ \cline{1-2}

$Y_t^{Q}$ &
Target function under $Q$-learning
\\ \cline{1-2}

$Y_t^{S}$ &
Target function under \ac{SARSA}
\\ \cline{1-2}

$\gamma$ &
Discount factor, where $0<\gamma<1$
\\ \cline{1-2}

$\alpha$ &
Learning rate
\\ \cline{1-2}

$\epsilon$ &
Exploration rate in an $\epsilon$-greedy policy
\\ \cline{1-2}

${Q}_{\boldsymbol\theta}(s,a)$ &
Linearly approximated action-value function with the parameter vector $\boldsymbol\theta$, used by the linear function approximation methods
\\ \cline{1-2}

\end{tabular}
\label{Tab:RLNotations}
\end{table}

\section{Problem Formulation of \texttt{SliceMapper}}
\label{Sec:ProblemFormulation}

We consider $h$ sequential \ac{VNFC} requests and $m$ candidate \acp{VM}. Request $f_i$ requires storage $S_i^{\rm req}$ and compute $C_i^{\rm req}$, whereas VM $v_j$ provides capacities $S_j^{\rm cap}$ and $C_j^{\rm cap}$. A \ac{VM} may host several \acp{VNFC}, given neither of the capacities is exceeded. Hence, the feasibility of each subsequent placement depends on earlier placement decisions.

Let $x_{ij}\in\{0,1\}$ indicate whether $f_i$ is mapped to $v_j$, let $y_j\in\{0,1\}$ indicate whether $v_j$ is used, and let $z_i\in\{0,1\}$ indicate whether $f_i$ is placed. The feasibility constraints are
\begin{align}
 \sum_{j=1}^{m}x_{ij} &= z_i,
 && i=1,\ldots,h,
 \label{eq:assignment-revised}\\
 \sum_{i=1}^{h}S_i^{\rm req}x_{ij}
     &\le S_j^{\rm cap}y_j,
 && j=1,\ldots,m,
 \label{eq:storage-revised}\\
 \sum_{i=1}^{h}C_i^{\rm req}x_{ij}
     &\le C_j^{\rm cap}y_j,
 && j=1,\ldots,m,
 \label{eq:compute-revised}\\
 x_{ij} &\textcolor{black}{\le y_j,}
 && \textcolor{black}{i=1,\ldots,h,\;j=1,\ldots,m,}
 \label{eq:vm-active-link-1}\\
 \textcolor{black}{y_j} &\textcolor{black}{\le \sum_{i=1}^{h}x_{ij},}
 && \textcolor{black}{j=1,\ldots,m.}
 \label{eq:vm-active-link-2}
\end{align}

For a feasible complete mapping, $z_i=1$, $\forall i\in\{1,\ldots,h\}$. Retaining $z_i$ in the optimization problem also permits a meaningful result when the instance is globally infeasible. Constraints~\eqref{eq:vm-active-link-1} and \eqref{eq:vm-active-link-2} ensure that $y_j=1$ exactly when $v_j$ is active. For a completed placement, define the residual storage and compute resources, respectively, as 
\begin{align}
r_j^S = r_{j,h}^S
&= S_j^{\rm cap}
   - \sum_{i=1}^{h} S_i^{\rm req} x_{ij}, \\
r_j^C = r_{j,h}^C
&= C_j^{\rm cap}
   - \sum_{i=1}^{h} C_i^{\rm req} x_{ij}.
\end{align}

Let $U=\sum_{j=1}^{m} y_j$, $M=\min\{h,m\}$. Define the normalized residual-resource function as
\begin{equation}
\begin{aligned}
W = \sum_{j=1}^{m} \Biggl[
y_j - \frac{1}{2} \Biggl(
& \frac{\sum_{i=1}^{h} S_i^{\rm req} x_{ij}}
       {S_j^{\rm cap}} \\
& {}+ \frac{\sum_{i=1}^{h} C_i^{\rm req} x_{ij}}
          {C_j^{\rm cap}}
\Biggr) \Biggr].
\end{aligned}
\label{eq:normalized-waste}
\end{equation}

The normalized efficiency cost is computed as \begin{equation}
 J_{\beta}=\frac{(1-\beta)U+\beta W}{M},
 \qquad 0\leq\beta\leq1,
 \label{eq:efficiency-cost}
\end{equation}
where a smaller value denotes a more consolidated placement with less normalized residual capacity. In all reported experiments, we set $\beta=0.5$, i.e., 
\[
     J_{\beta} = \frac{U + W}{2M}.
\] Since storage units and compute units are not measured using a common measurement unit, this normalization is important as it converts these quantities into dimensionless fractions. Experiments consider two residual wastage summaries, namely 
\begin{equation}
\label{eq:W1}
    W_{\rm norm} = W/M,
\end{equation}
and 
\begin{equation}
\label{eq:W2}
    W_{\rm act} = \begin{cases}
        W/U,
&
\text{if } U>0,
\\[1.5mm]
1,
&
\text{if } U = 0,
    \end{cases}
\end{equation}
where $W_{\rm norm}$ is the normalized residual wastage and $W_{\rm act}$ denotes the mean residual resource fraction over active \acp{VM}.

The optimization strictly prioritizes the number of successfully placed \acp{VNFC}, while using $J_\beta$ as a secondary resource efficiency criterion. In the implemented \ac{MILP} baseline, this priority is realized within a single \ac{MILP}
\begin{equation}
\label{eq:milp-objective}
\min_{x,y,z}
\quad
J_\beta
-
C\sum_{i=1}^{h}z_i,
\end{equation}
subject to the constraints in Equations~\eqref{eq:assignment-revised} - \eqref{eq:vm-active-link-2}, and $C = 2$ denotes the dominance coefficient.

Since $0 \leq J_{\beta} \leq 1$, setting $C=2$ ensures that placing one additional \ac{VNFC} always improves the objective more than any possible improvement in the normalized efficiency cost. The resulting \ac{MILP} maximizes the number of successful placements. If two or more solutions have the same placement count, it favors the one with the lower efficiency cost. Expanding the terms in Equation (\ref{eq:milp-objective}) linearly, the objective can equivalently be written as
\begin{align}
\min_{x,y,z}\quad
&
\frac{1}{M}\sum_{j=1}^{m}y_j
-
\frac{\beta}{2M}
\sum_{i=1}^{h}\sum_{j=1}^{m}
x_{ij}
\left(
\frac{S_i^{\rm req}}{S_j^{\rm cap}}
+
\frac{C_i^{\rm req}}{C_j^{\rm cap}}
\right)
\nonumber\\
&
-
C\sum_{i=1}^{h}z_i.
\label{eq:milp-objective-expanded}
\end{align}

The \ac{MILP} baseline is solved using SciPy's \texttt{milp}. A relative \ac{MILP}-gap tolerance of $10^{-6}$ and a time limit of $30$s are set. Solver status, runtime, best bound, and relative gap are retained. When optimality is established, results are reported as solver-certified; otherwise, as time-bounded incumbents.


\section{Proposed Solution for \texttt{SliceMapper}}
\label{Sec:ProposedSolution}
Following the system model and capacity-constrained placement formulation introduced in the preceding sections, we now describe the learning-assisted solution used by \texttt{SliceMapper}. The placement process is treated as a finite sequential decision-making problem in which the \ac{VNFC} requests are processed in a fixed order and the residual resources of the candidate \acp{VM} evolve after each successful placement. The following four temporal-difference implementations are considered: tabular \ac{SARSA}, tabular $Q$-learning, linear \ac{SARSA}, and linear $Q$-learning\footnote{Linear denotes the linear functionals under usage in the function approximation method.}. These are well-established \ac{RL} algorithms, which will be used to solve the \ac{VNFC}-to-\ac{VM} placement problem defined in Sections \ref{Sec:SystemModel} and \ref{Sec:ProblemFormulation}. All four methods interact with the same feasibility-aware environment.

\subsection{Overview of the RL-Based Solution for VNFC Mapping}
\label{Subsec:RLOverview}

The agent, i.e., \texttt{SliceMapper}\footnote{The terms \emph{agent} and \texttt{SliceMapper} are used interchangeably.}, interacts with the placement environment by observing its current state, selecting a feasible \ac{VM} action, receiving a reward, and proceeding through the ordered \ac{VNFC} request sequence. Let $t\in\{0,\ldots,h-1\}$. At time step $t$, the resource demand vector, defined as
\begin{equation}
d_t
=
(S_{t+1}^{\rm req},C_{t+1}^{\rm req})
\label{eq:demand-code-index}
\end{equation}
stores the storage and compute demands of the \ac{VNFC} $f_{t+1}$. A placement step $t$ is a single decision within an episode, whereas an episode indicates a complete attempt to process the entire batch of all the \acp{VNFC}. The complete ordered request batch is assumed to be known at the
beginning of a placement step $t$ in an episode. The environmental state is given by
\begin{equation}
S_t
=
\left(
t,d_t,
\left\{
S_j^{\rm cap},C_j^{\rm cap},
r_{j,t}^{S},r_{j,t}^{C}
\right\}_{j=1}^{m}
\right),
\label{eq:formal-state}
\end{equation}
where the quantities $r_{j,t}^{S},r_{j,t}^{C}$ denote the storage and compute residual capacities of \ac{VM} $v_j$ at placement step $t$, respectively. Environmental states record the current request, the storage and compute capacities of the candidate \acp{VM}, and their residual resources remaining after all preceding placements.

For each current request, an action corresponds to selecting one candidate \ac{VM}. The action space $\mathcal{A} = \{1,\ldots,m\}$ contains all the $m$ candidate \ac{VM} indices, but one restricts to the feasible subset
\begin{equation}
\mathcal A_f(S_t)
=
\left\{
j\in\{1,\ldots,m\} \mid
S_{t+1}^{\rm req}\leq r_{j,t}^{S},
\;
C_{t+1}^{\rm req}\leq r_{j,t}^{C}
\right\}.
\label{eq:feasible-action-set}
\end{equation} Consequently, infeasible \ac{VM} actions are not presented to the learning policy. If \ac{VM} $j\in\mathcal A_f(S_t)$ is selected, its residual resources are updated via
\begin{equation}
r_{j,t+1}^{S}
=
r_{j,t}^{S}-S_{t+1}^{\rm req},
\qquad
r_{j,t+1}^{C}
=
r_{j,t}^{C}-C_{t+1}^{\rm req},
\label{eq:rl-residual-update}
\end{equation}
while the residual resources of every other \ac{VM} remain unchanged. A \ac{VM} may host several \acp{VNFC} provided that sufficient storage and compute resources are available. Initially, every \ac{VM} $j$ has 
\[
    r_{j,0}^{S} = S_j^{\rm cap}, ~~ r_{j,0}^{C} = C_j^{\rm cap}.
\]

If $\mathcal A_f(S_t)=\varnothing$, then no action is feasible, and the agent has no decision to make. The environment performs rejections as follows: the request at time step $t$ is marked unplaced, a reward of $-1$ is given, residual resources remain unaltered, and
the request index advances to time step $t+1$. Process proceeds further until all $h$ \ac{VNFC} requests have been considered.

The reward at time step $t + 1$ is given by
\begin{equation}
R_{t+1}
=
\begin{cases}
-1,
&
\mathcal A_f(S_t)=\varnothing,
\\[1.5mm]
\dfrac{1}{h}
+
i_t\eta(1-J_\beta),
&
A_t\in\mathcal A_f(S_t),
\end{cases}
\label{eq:revised-reward}
\end{equation}
where
\begin{equation}
\eta=0.25,
\qquad
\beta=0.5,
\end{equation}
and 
\begin{equation}
i_t
=
\begin{cases}
1,
&
\text{if } t=h-1
\text{ and preceding \acp{VNFC} have been placed},
\\[1.5mm]
0,
&
\text{otherwise}.
\end{cases}
\label{eq:completion-indicator}
\end{equation} 
Each successful placement is therefore destined to contribute at least a reward $1/h$ (equality occurs when $i_t = 0$). When $i_t = 1$, i.e., when all $h$ \acp{VNFC} are placed successfully, we have an additional efficiency bonus of $\eta(1-J_\beta)$. We know that $0\leq J_\beta\leq1$. Therefore, this efficiency bonus lies in the interval $[0,0.25]$.


\subsection{Action Value and Placement Policy}
\label{Subsec:ActionValue}
A policy $\pi(a|s)$ specifies the probability of selecting action $a$ in state $s$. For a finite placement episode, the discounted return from step $t$ is
\begin{equation}
G_t
=
\sum_{k=0}^{T-t-1}
\gamma^k R_{t+k+1},
\label{eq:finite-return}
\end{equation}
where $\gamma\in(0,1)$ is the discount factor and $T$ denotes the end of the episode. The action-value function associated with policy $\pi$ is
\begin{equation}
Q_\pi(s,a)
=
\mathbb E_\pi
\left[
G_t
\mid
S_t=s,\;
A_t=a
\right].
\label{eq:q-policy}
\end{equation} It represents the expected discounted return obtained after selecting the action $a$ in state $s$ and subsequently following policy $\pi$. During training, action selection is $\epsilon$-greedy over $\mathcal A_f(S_t)$ only, and after training, the learned action values are evaluated using a greedy policy.

Let $k(t)$ be the number of consecutive forced rejections taking place after action $A_t$. The rewards are accumulated as
\begin{equation}
\overline{R(t)}
=
R_{t+1}
+
\sum_{\ell=1}^{k(t)}
\gamma^\ell R_{t+\ell+1}.
\label{eq:accumulated-reward}
\end{equation}
For instance, if $k(t) = 1$, then 
\[
    \overline{R(t)}
=
R_{t+1}
+ \gamma R_{t+2} = R_{t+1} - \gamma. 
\]
Note that when $k(t)=0$, the formulation reduces to the standard one-step temporal-difference target.


\subsection{On-Policy SARSA and Off-Policy $Q$-Learning} \label{Subsec:QalgoGeneral}
In \ac{SARSA}, the next targeted action is generated by the same $\epsilon$-greedy behavior policy used in interaction with the environment, which makes it an on-policy \ac{RL} technique. In $Q$-learning, the behavior policy may remain exploratory while the target uses the largest estimated value among feasible next actions. Hence, $Q$-learning is off-policy.

The tabular \ac{SARSA} target is given by
\begin{equation}
Y_t^{S}
=
\overline{R(t)}
+
\gamma^{k(t)+1}
Q(s',a'),
\label{eq:tabular-sarsa-target}
\end{equation}
where the action $a'$ is selected at state $s'$ by the same $\epsilon$-greedy behavior policy. The tabular $Q$-learning target is given by
\begin{equation}
Y_t^{Q}
=
\overline{R(t)}
+
\gamma^{k(t)+1}
\max_{a\in\mathcal A_f(s')}
Q(s',a).
\label{eq:tabular-q-target}
\end{equation}
The tabular value update appearing in both methods is given by \begin{equation}
Q(S_t,A_t)
\leftarrow
Q(S_t,A_t)
+
\alpha_n
\left[
Y_t-Q(S_t,A_t)
\right],
\label{eq:tabular-common-update}
\end{equation}
where $\alpha_n$ denotes the learning rate at a corresponding update.


\begin{algorithm}[t]
\caption{Tabular SARSA}
\label{alg:tab-sarsa}
\begin{algorithmic}[1]
\footnotesize

\Require 
\begin{itemize}
    \item Ordered \ac{VNFC} batch
    \item \ac{VM} capacities
    \item  $0 < \gamma < 1$; $0 \leq \epsilon \leq 1$; $0 \leq \alpha_n \leq1$
\end{itemize}

\State Initialize $Q(\tilde s,a) = 0$

\For{each episode}
    \State Reset environment advance through the forced rejections and obtain the first decision state $S_t$
    \While{a decision state exists}

        \State Construct $\tilde S_t$
        \State Select $\epsilon$-greedy feasible action $A_t$

        \State Execute $A_t$, advancing through any forced rejections, obtaining the $3$-tuple $(\overline{R(t)},k(t),s')$

        \If{$s'$ happens to be the terminal state}
            \State $Y_t^{S}=\overline{R(t)}$
        \Else
            \State Select $\tilde{s}'$ and $a'$ by the same $\epsilon$-greedy policy
            \State
            Set $Y_t^{S} = 
            \overline{R(t)}+\gamma^{k(t)+1}Q(\tilde s',a')$
        \EndIf

        \State
        $Q(\tilde S_t,A_t) \leftarrow
        Q(\tilde S_t,A_t)+
        \alpha_n[Y_t^{S}-Q(\tilde S_t,A_t)]$
        \If{$s'$ happens to be non-terminal state}
        \State $(S_t,\tilde{S_t},A_t)=(s', \tilde{s}' ,a')$
        \EndIf

    \EndWhile
\EndFor

\end{algorithmic}
\end{algorithm}

\begin{algorithm}[t]
\caption{Tabular $Q$-Learning}
\label{alg:tab-qlearning}
\begin{algorithmic}[1]
\footnotesize

\Require
\begin{itemize}
    \item Ordered \ac{VNFC} batch
    \item \ac{VM} capacities
    \item  $0 < \gamma < 1$; $0 \leq \epsilon \leq 1$; $0 \leq \alpha_n \leq1$
\end{itemize}

\State Initialize $Q(\tilde s,a)=0$

\For{each episode}
    \State Reset environment advance through the forced rejections and obtain the first decision state $S_t$
    \While{a decision state exists}

        \State Construct $\tilde S_t$
        \State Select $\epsilon$-greedy feasible action $A_t$

        \State Execute $A_t$, advancing through any forced rejections, obtaining the $3$-tuple $(\overline{R(t)},k(t),s')$

        \If{$s'$ happens to be the terminal state}
            \State $Y_t^{Q}=\overline{R(t)}$
        \Else
            \State Set
            $Y_t^{Q}=
            \overline{R(t)}+
            \gamma^{k(t)+1}
            \max_{a\in\mathcal A_f(s')}
            Q(\tilde s',a)$
        \EndIf

        \State
        $Q(\tilde S_t,A_t) \leftarrow
        Q(\tilde S_t,A_t)+
        \alpha_n[Y_t^{Q}-Q(\tilde S_t,A_t)]$

        \State $S_t= s'$

    \EndWhile
\EndFor

\end{algorithmic}
\end{algorithm}

\subsection{Tabular SARSA and Tabular $Q$-Learning} \label{Subsec:TabularMethods}
The tabular methods explicitly store the value of encountered state-action pairs. Because the continuous resource quantities must be represented by a finite key, each normalized component is discretized using
\begin{equation}
q_L(x)
=
\min
\left\{
L-1,\;
\left\lfloor
L\,\chi(x;[0,1])
\right\rfloor
\right\},
\qquad
L=5.
\label{eq:tabular-quantizer}
\end{equation}

where 
\[
    \chi(x;[0,1]) = \begin{cases}
        0,
&
x < 0,
\\[1.5mm]
x,
&
0\leq x \leq 1,
\\[1.5mm]
1, & x > 1
    \end{cases}
\]
is a function supported in the interval $[0,1]$.
Let $\widetilde S_t$ denote the resulting tabular key. It is a tuple consisting of the request position $t$, the normalized current storage, the normalized compute demand, the normalized capacity, and the normalized residual-resource coordinates of all candidate \acp{VM}. The normalization of storage and compute capacities happens with respect to maximum storage and compute capacities, $S_{\rm max}$ and $C_{\rm max}$, respectively. The normalization of storage and compute residuals happens with respect to $S_j^{\rm cap}$ and $C_j^{\rm cap}$, respectively.

\texttt{Algorithm~\ref{alg:tab-sarsa}} and \texttt{Algorithm~\ref{alg:tab-qlearning}} summarize the two tabular learning methods. The pseudocodes present the defining learning logic. One of the main advantages of the tabular representation methods is that distinct values can be retained for individual encountered states, whereas its primary limitation is representation growth with increasing candidate \acp{VM}. The implementation allows only the encountered state keys to be retained and does not study the complete state space. Each encountered key stores an action-value vector of length $m$.

\subsection{Linear SARSA and Linear $Q$-Learning} \label{Subsec:LinearMethods}
The linear methods replace the explicit tabular representation with a shared parameterized approximation
\begin{equation}
Q_{\boldsymbol\theta}(S_t,j)
=
\boldsymbol\theta^{\mathsf T}
\boldsymbol\phi(S_t,j),
\label{eq:linear-q}
\end{equation}
where
\begin{equation}
\boldsymbol\phi(S_t,j)\in\mathbb R^{15},
\qquad
\boldsymbol\theta\in\mathbb R^{15}
\end{equation}
are $15$-dimensional feature and parameter vectors, respectively.

The following $15$ features were used to summarize the placement step and current request demand; \textit{the selected \acp{VM} residual storage resources before a hypothetical placement, the selected \acp{VM} residual storage resources after a hypothetical placement, the selected \acp{VM} residual compute resources before a hypothetical placement, the selected \acp{VM} residual compute resources after a hypothetical placement, feasibility, status about the current \ac{VM}-activation, post-placement fit, storage-compute balance, remaining storage demand pressure, remaining compute demand pressure, the current fraction of active \acp{VM}, normalized current \ac{VNFC} storage demand, normalized current \ac{VNFC} compute demand, the placement progress term $z_t^{h}:= t/h$, and a constant bias term $1$}. Note that $z_t^{h}$ tells us about how far the placement scenario across the ordered \ac{VNFC} batch has progressed. For instance, if $z_t^{h} = 0$, then we are at the beginning of the placement scenario, whereas when $z_t^{h} =\frac{h-1}{h}$, then we are at the final \ac{VNFC} placement decision. The complete ordered request batch is known at the start of the episode.

For a feasible placement, the post-placement fit feature is given by $1 - 1/2 \times$\{normalized post-placement storage and compute residuals\}. The compute-storage balance feature is given by the absolute difference between these two residual fractions subtracted from $1$. Therefore, deviations in either direction represent a symmetric imbalance. For both linear methods, the parameter update is
\begin{equation}
\boldsymbol\theta
\leftarrow
\boldsymbol\theta
+
\alpha_n
\left[
\left(
Y_t-
Q_{\boldsymbol\theta}(S_t,A_t)
\right)
\boldsymbol\phi(S_t,A_t)
-
\lambda\boldsymbol\theta
\right],
\label{eq:linear-common-update}
\end{equation}
where $\lambda$ is the $\ell_2$ regularization coefficient. For linear \ac{SARSA}, we have
\begin{equation}
Y_t^{S}
=
\overline{R(t)}
+
\gamma^{k(t)+1}
Q_{\boldsymbol\theta}(s',a'),
\label{eq:linear-sarsa-target}
\end{equation}
where $a'$ is selected by the same $\epsilon$-greedy behavior policy. For linear $Q$-learning, we have 
\begin{equation}
Y_t^{Q}
=
\overline{R(t)}
+
\gamma^{k(t)+1}
\max_{a\in\mathcal A_f(s')}
Q_{\boldsymbol\theta}(s',a).
\label{eq:linear-q-target}
\end{equation}

\begin{algorithm}[t]
\caption{Linear SARSA}
\label{alg:linear-sarsa}
\begin{algorithmic}[1]
\footnotesize

\Require \begin{itemize}
    \item Ordered \ac{VNFC} batch
    \item \ac{VM} capacities
    \item  $0 < \gamma < 1$; $0 \leq \epsilon \leq 1$; $0 \leq \alpha_n \leq1$
    \item the $\ell_2$ regularizer $\lambda$
\end{itemize}

\State Initialize the vector $\boldsymbol\theta=\mathbf 0$

\For{each episode}
    \State Reset environment advance through the forced rejections and obtain the first decision state $S_t$
    \While{a decision state exists}

        \State Select $A_t$ $\epsilon$-greedily using
        $Q_{\boldsymbol\theta}(S_t,a)$

        \State Execute $A_t$, advancing through any forced rejections, obtaining the $3$-tuple $(\overline{R(t)},k(t),s')$

        \If{$s'$ is terminal}
            \State $Y_t^{S}=\overline{R(t)}$
        \Else
            \State Select $a'$ by the same $\epsilon$-greedy policy
            \State
            Set $Y_t^{S}=
            \overline{R(t)}+
            \gamma^{k(t)+1}
            Q_{\boldsymbol\theta}(s',a')$
        \EndIf

        \State
        $\boldsymbol\theta \leftarrow
        \boldsymbol\theta+
        \alpha_n
        \big[
        (Y_t^{S}-Q_{\boldsymbol\theta}(S_t,A_t))
        \boldsymbol\phi(S_t,A_t)
        -
        \lambda\boldsymbol\theta
        \big]$
        \If{$s'$ happens to be non-terminal state}
        \State $(S_t,A_t)=(s',a')$
        \EndIf

    \EndWhile
\EndFor

\end{algorithmic}
\end{algorithm}

\begin{algorithm}[t]
\caption{\color{black} Linear $Q$-Learning}
\label{alg:linear-qlearning}
\begin{algorithmic}[1]
\footnotesize

\Require  \begin{itemize}
    \item Ordered \ac{VNFC} batch
    \item \ac{VM} capacities
    \item  $0 < \gamma < 1$; $0 \leq \epsilon \leq 1$; $0 \leq \alpha_n \leq1$
    \item the $\ell_2$ regularizer $\lambda$
\end{itemize}

\State Initialize the vector $\boldsymbol\theta=\mathbf 0$

\For{each episode}
    \State Reset environment advance through the forced rejections and obtain the first decision state $S_t$
    \While{a decision state exists}

        \State Select $A_t$ $\epsilon$-greedily using
        $Q_{\boldsymbol\theta}(S_t,a)$

        \State Execute $A_t$, advancing through any forced rejections, obtaining the $3$-tuple $(\overline{R(t)},k(t),s')$

        \If{$s'$ is terminal}
            \State $Y_t^{Q}=\overline{R(t)}$
        \Else
            \State
            Set $Y_t^{Q}=
            \overline{R(t)}+
            \gamma^{k(t)+1}
            \max_{a\in\mathcal A_f(s')}
            Q_{\boldsymbol\theta}(s',a)$
        \EndIf

        \State
        $\boldsymbol\theta \leftarrow
        \boldsymbol\theta+
        \alpha_n
        \big[
        (Y_t^{Q}-Q_{\boldsymbol\theta}(S_t,A_t))
        \boldsymbol\phi(S_t,A_t)
        -
        \lambda\boldsymbol\theta
        \big]$

        \State $S_t= s'$

    \EndWhile
\EndFor
\end{algorithmic}
\end{algorithm}

\texttt{Algorithm~\ref{alg:linear-sarsa}} and \texttt{Algorithm~\ref{alg:linear-qlearning}} summarize the two linear implementations. The linear representation uses only $15$ learned features. These features are independent of the number of \acp{VM}. This significantly helps in the reduction of model storage when compared with the tabular representation counterparts. However, each feasible \ac{VM} must still be evaluated through $Q_{\boldsymbol\theta}(S_t,j)$, so a fixed parameter dimension does not guarantee the generalization of the methods to arbitrary problems with arbitrary sizes.

\section{Performance Evaluation of \texttt{SliceMapper}} \label{Sec:PerformanceEvaluationRevisedFinal}
This section provides an empirical study of the proposed \texttt{SliceMapper} agent through a controlled set of experiments. The four temporal-difference methods, i.e., tabular \ac{SARSA}, tabular $Q$-learning, linear \ac{SARSA}, and linear $Q$-learning, are compared with the baselines considered in this article, namely random-feasible placement, first-fit, first-fit-decreasing, best-fit, resource-wastage-aware heuristic, and \ac{MILP}. The evaluations are based on the placed-\ac{VNFC} fraction, placement-efficiency cost, active-\ac{VM} residual wastage, online decision time, computational cost of training, and representation memory. Additionally, the empirical stabilization, scalability across controlled problem sizes, behavior under failure and distribution-shift scenarios, and the effect of terminal efficiency shaping are also examined.

\subsection{Experimental Setup, Baselines, and Reported Metrics} \label{Subsec:ExperimentalProtocol}
We compare four \ac{RL} methods with random feasible placement, first-fit, first-fit decreasing, best-fit, a resource-wastage-aware heuristic, and the \ac{MILP} baseline described by Equation~\eqref{eq:milp-objective}. The final experiments were conducted using the paper-profile configuration summarized in Table~\ref{Tab:ExperimentalConfiguration}. The maximum training budget is $3000$ episodes. The empirical stabilization test is enabled only after a minimum of $750$ episodes. Greedy-policy evaluation is performed every $50$ episodes using the stabilization criterion described in Subsection \ref{Subsec:LearningStabilization}.

Hyperparameters are selected exclusively using seeds $100$-$104$. For every controlled problem size and each benchmark seed in $300$-$329$, a fresh \ac{RL} model is trained on a generated placement instance, and its final greedy policy is evaluated on that same instance. Thus, the main benchmark measures instance-specific optimization performance after learning and is not a held-out or zero-shot generalization test. The separate distribution-shift experiment described in Subsection~\ref{Subsec:FailureDistributionShift} is the only evaluation in which a trained policy is applied without retraining to modified instances.

Random-feasible placement samples uniformly from the feasible action indices, while first-fit selects the first feasible \ac{VM} in the fixed candidate order. First-fit decreasing first sorts requests, stably and in descending order, by
\begin{equation}
d_i^{\rm dom}=\max\!\left\{
\frac{S_i^{\rm req}}{S_{\rm max}},
\frac{C_i^{\rm req}}{C_{\rm max}}
\right\},
\label{eq:ffd-order}
\end{equation}
and then applies first-fit. For a feasible candidate $j$, the best fit minimizes
\begin{equation}
b_{ij}=\frac{1}{2}\!\left(
\frac{r_{j,t}^{S}-S_i^{\rm req}}{S_j^{\rm cap}}+
\frac{r_{j,t}^{C}-C_i^{\rm req}}{C_j^{\rm cap}}
\right),
\label{eq:best-fit-score}
\end{equation}
whereas the resource-wastage-aware heuristic minimizes
\begin{equation}
b_{ij}+\left|
\frac{r_{j,t}^{S}-S_i^{\rm req}}{S_j^{\rm cap}}-
\frac{r_{j,t}^{C}-C_i^{\rm req}}{C_j^{\rm cap}}
\right|+0.05\,\mathbf 1\{v_j\text{ is inactive}\}.
\label{eq:rwa-score}
\end{equation}
First-fit decreasing and \ac{MILP} are batch-aware. Random feasible, first-fit, best-fit, resource-wastage-aware, and four learned temporal-difference methods process requests sequentially.

The quantities $S_j^{\rm cap}$ and $C_j^{\rm cap}$ were sampled independently from the discrete uniform distributions
$\mathcal{DU}\{80,\ldots,180\}$ and
$\mathcal{DU}\{8,\ldots,20\}$, respectively. On the other hand, the \ac{VNFC} requirements, i.e., $S_i^{\rm  req}$ and $C_i^{\rm  req}$ were generated from independent uniform random weights $\mathcal{U}(0.5,1.5)$, rescaled separately for each resource to a target aggregate load of $0.75$, capped at $90\%$ of the corresponding maximum \ac{VM} capacity, which is then rounded to positive integers. Moreover, all the quantities are expressed in synthetic resource units. The realized aggregate storage and compute loads are
\begin{equation}
l_S=\frac{\sum_{i=1}^{h}S_i^{\rm req}}{\sum_{j=1}^{m}S_j^{\rm cap}},
\qquad
l_C=\frac{\sum_{i=1}^{h}C_i^{\rm req}}{\sum_{j=1}^{m}C_j^{\rm cap}}.
\label{eq:aggregate-load}
\end{equation}
The value $0.75$ is the pre-clipping target, and the realized values after the per-request capping and integer rounding may slightly differ. The controlled-load scalability \ac{VNFC}-to-\ac{VM} scenarios considered are \[(h,m)\in \{(8,4),(16,8),(32,16),(64,32)\}, \] with a target aggregate load of $0.75$ for both compute and storage resources.

The final evaluation uses $30$ independent test instances per controlled problem size; hyperparameter selection uses $5$ independent tuning instances; the failure and distribution-shift analysis uses $10$ independent instances; and the reward-ablation study uses $10$ independent instances. For each method and problem size, we report the placed-\ac{VNFC} fraction, $J_\beta$, the number of active \acp{VM}, normalized residual wastage, decision time per \ac{VNFC}, and, if applicable, training time and representation cost. For tabular methods, we additionally report visited states and stored state-action values; for the linear methods, the learned representation contains a fixed $15$-dimensional parameter vector.

For an evaluation set containing $N$ independent instances, let $z_i^{(k)}$ indicate whether a request $i$ is placed in instance $k$. The reported placement metric is
\begin{equation}
c_{\rm place}^{(k)}=\frac{1}{h}\sum_{i=1}^{h}z_i^{(k)},
\qquad
C_{\rm place}=\frac{1}{N}\sum_{k=1}^{N}c_{\rm place}^{(k)}.
\label{eq:placed-vnfc-fraction}
\end{equation}
Thus, $C_{\rm place}$ is the mean fraction of placed requests, not the probability that an entire instance is completed. For the principal placement-quality metrics, the reported mean is accompanied by a normal-approximation $95\%$ confidence interval computed across independent seeds as \begin{equation}
\bar{x}
\pm
1.96\,\frac{s}{\sqrt{n}},
\label{eq:normal-ci}
\end{equation}
where $\bar{x}$ is the sample mean, $s$ is the sample standard deviation, and $n$ is the number of independent instances.

Training-time distributions may be strongly skewed, particularly at larger problem sizes. Hence, it is summarized using the median and inter-quartile range. The conventional heuristics apply fixed placement rules and require no offline policy-training stage. The four temporal-difference methods incur offline training costs before greedy-policy evaluation. For learned policies, decision timing includes state encoding or feature construction and greedy action selection. For heuristics, it includes the placement chooser but excludes environment updates. First-fit-decreasing preprocessing and sorting are excluded. For \ac{MILP}, the timer covers the complete solver call and is divided by $h$ when reported per \ac{VNFC}. Thus, the measurements compare the implemented online decision paths but do not have identical low-level timing boundaries. All methods within a configuration are measured on the same host, so the timing results support within-study comparisons rather than platform-independent latency guarantees. The \ac{MILP} solver is subject to a $30$ seconds time limit. Solver status and relative \ac{MILP} gap are retained so that solver-certified solutions can be distinguished from time-bounded incumbents.

\begin{table}[t]
\caption{Experimental setup used in the final evaluation}
\label{Tab:ExperimentalConfiguration}
\centering
\renewcommand{\arraystretch}{1.40}
\setlength{\tabcolsep}{2.5pt}
\resizebox{\columnwidth}{!}{%
\begin{tabular}{|l|c|}
\hline
\rowcolor{tableheader}
\multicolumn{1}{|c|}{\textcolor{white}{\textbf{Parameter}}}
&
\multicolumn{1}{c|}{\textcolor{white}{\textbf{Final Setting}}}
\\
\hline
Controlled problem sizes &
$8\times4$, $16\times8$, $32\times16$, $64\times32$
\\
\hline
Target aggregate load & $0.75$ per resource \\
\hline
Maximum training episodes & $3000$ \\
\hline
Minimum episodes before stabilization test & $750$ \\
\hline
Greedy evaluation interval & $50$ episodes \\
\hline
Stability window & $5$ checkpoints \\
\hline
Test seeds & $30$ independent seeds ($300-329$)\\
\hline
Tuning seeds & $5$ independent seeds ($100-104$) \\
\hline
Failure/shift seeds & $10$ independent seeds ($500-509$) \\
\hline
Reward-ablation seeds & $10$ independent seeds ($700-709$)\\
\hline
\ac{MILP} time limit & $30$ s \\
\hline
\ac{MILP} relative gap tolerance & $10^{-6}$ \\
\hline
Tabular discretization levels & $L=5$ \\
\hline
Linear parameter dimension & $15$ \\
\hline
\end{tabular}%
}
\end{table}

\subsection{Hyperparameter Selection}\label{Subsec:HyperparameterSelection}
Four learning methods are tuned independently over eight candidate
hyperparameter configurations using the same five tuning instances. Selected configurations are reported in Table~\ref{Tab:SelectedHyperparameters}. This method-specific selection is necessary because the tabular and linear representations exhibit different update dynamics and hence need not share identical learning or exploration parameters. Selected configurations use constant learning rates throughout training, i.e., $\texttt{alpha\_power}=0$ for all four methods. The two linear methods also employ an $\ell_2$ regularization coefficient of $10^{-5}$. No regularization term is used by the tabular methods. The hyperparameter $d_{\epsilon}$ denotes the maximum training budget fraction over which $\epsilon$ decays from $\epsilon_{\mathrm{start}}$ to $\epsilon_{\mathrm{end}}$. Based on the lower divergence rate, higher mean placed fraction, higher mean evaluation score, lower mean $J_\beta$, and lower mean training time, the candidates are ordered lexicographically. Benchmark, failure-analysis, and ablation seeds are not used for hyperparameter selection.

\subsection{Learning Behavior and Empirical Stabilization}\label{Subsec:LearningStabilization}
The learning methods are trained for at most $N_{\max}=3000$ episodes. To avoid equating a flat learning curve with theoretical convergence, we use an operational empirical-stabilization criterion based on periodic greedy-policy evaluation. Every $K=50$ training episodes, the current greedy policy is evaluated without exploration. Let $C_e$ denote the placed-\ac{VNFC} fraction obtained at evaluation checkpoint $e$, and let $J_{\beta,e}$ denote the corresponding efficiency cost. The checkpoint score is defined as
\begin{equation}
\label{eq:stability-score}
G_e
=
C_e
+
\eta
\mathbf{1}\{C_e=1\}
\left(1-J_{\beta,e}\right),
\end{equation}
where $\eta=0.25$ is the same efficiency-weight coefficient used in the reward definition.

A run is declared empirically stable only after at least $750$ training episodes and when the latest $P=5$ evaluation checkpoints satisfy
\begin{equation}
C_e \geq 0.999
\end{equation}
for every checkpoint in the window and
\begin{equation}
\max G_e-\min G_e \leq 0.01
\end{equation}
over that same window. The reported stabilization episode corresponds to first episode in five-checkpoint window satisfying these conditions. Because the criterion is tested only after episode $750$, a reported value of $550$ means that checkpoints $550$, $600$, $650$, $700$, and $750$ have all combinedly passed the test. It does not mean that training stopped at episode $550$. Runs that do not satisfy the criterion continue until the hard cap of $3000$ episodes. This is a stopping rule and does not imply convergence to an optimal policy.

\begin{figure*}[t]
\begin{tcolorbox}
\centering
\includegraphics[width=0.98\textwidth]{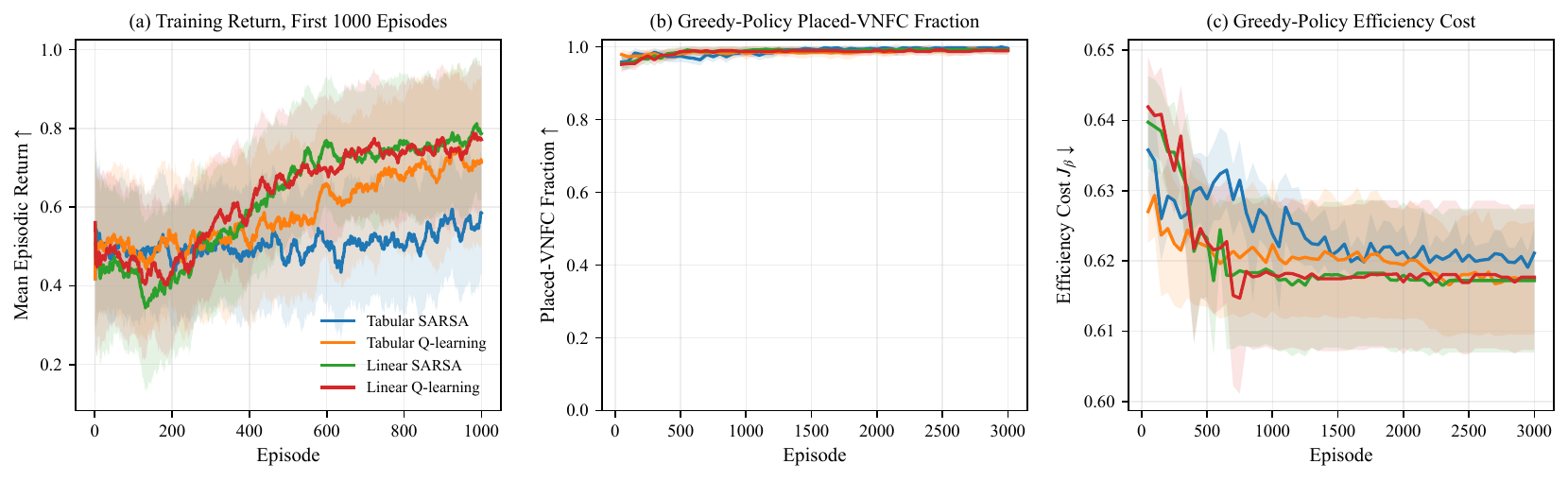}
\caption{Learning behavior of the four \ac{RL} methods:
(a) Mean episodic return over the first $1000$ training episodes,
(b) Periodically evaluated greedy-policy placed-\ac{VNFC} fraction, and
(c) Greedy-policy efficiency cost $J_\beta$. Curves in Plots~(b)--(c) are means over the $30$ independent reference instances and shaded bands are normal-approximation $95\%$ confidence intervals. Plot (a) is retained only as a comparison of the $4$ temporal-difference methods and shall not be used for statistical comparison or the stabilization decision.}
\label{Fig:LearningBehavior}
\end{tcolorbox}
\vspace{-4.8mm}
\end{figure*}

Figure~\ref{Fig:LearningBehavior} summarizes representative learning behavior. Panel~(a) reports mean episodic return over the first $1000$ episodes, whereas Panels~(b) and~(c) report periodically evaluated greedy-policy placed-VNFC fraction and efficiency cost, respectively. For runs that terminate early under the empirical stabilization criterion, the final observed greedy-policy checkpoint is carried forward in the visualization only and does not represent additional training.

\begin{table}[t]
\color{black}
\caption{Hyperparameters used when performing four learning methods}
\label{Tab:SelectedHyperparameters}
\centering
\scriptsize
\renewcommand{\arraystretch}{1.40}
\setlength{\tabcolsep}{2.0pt}
\begin{tabular}{|l|c|c|c|c|c|c|}
\hline
\rowcolor{tableheader}
\multicolumn{1}{|c|}{\textcolor{white}{\textbf{Method}}}
&
\multicolumn{1}{c|}{\textcolor{white}{$\boldsymbol{\alpha}$}}
&
\multicolumn{1}{c|}{\textcolor{white}{$\boldsymbol{\gamma}$}}
&
\multicolumn{1}{c|}{\textcolor{white}{$\boldsymbol{\epsilon_{\rm start}}$}}
&
\multicolumn{1}{c|}{\textcolor{white}{$\boldsymbol{\epsilon_{\rm end}}$}}
&
\multicolumn{1}{c|}{\textcolor{white}{$\boldsymbol{d_\epsilon}$}}
&
\multicolumn{1}{c|}{\textcolor{white}{$\boldsymbol{\lambda}$}}
\\
\hline
Tabular \ac{SARSA}
& $0.5$
& $0.9$
& $1.0$
& $0.05$
& $0.7$
& -- \\
\hline
Tabular $Q$-learning
& $0.1$
& $0.9$
& $0.5$
& $0.02$
& $0.5$
& -- \\
\hline
Linear \ac{SARSA}
& $0.005$
& $0.95$
& $0.7$
& $0.02$
& $0.7$
& $10^{-5}$ \\
\hline
Linear $Q$-learning
& $0.005$
& $0.95$
& $0.7$
& $0.02$
& $0.7$
& $10^{-5}$ \\
\hline
\end{tabular}
\end{table}

Table~\ref{Tab:EmpiricalStabilization} summarizes the empirical stabilization behavior across the $30$ independent test instances of the reference $16\times8$ configuration. The tabular \ac{SARSA} method satisfies the criterion in $29$ of $30$ runs, followed by tabular $Q$-learning in $28$ runs, linear \ac{SARSA} in $27$ runs, and linear $Q$-learning in $26$ runs. Among the stabilized runs, the median stabilization episode is $850$ for tabular \ac{SARSA} and $550$ for other three methods. The table is therefore used only to characterize the observed training behavior under the selected experimental settings, and the lower median stabilization episode of the two linear methods and tabular $Q$-learning should not be interpreted as evidence of universally faster or better learning.

\begin{table}[t]
\caption{Empirical stabilization behavior for the $16\times8$ configuration}
\label{Tab:EmpiricalStabilization}
\centering
\scriptsize
\renewcommand{\arraystretch}{1.40}
\setlength{\tabcolsep}{2.5pt}
\begin{tabular}{|l|c|c|c|}
\hline
\rowcolor{tableheader}
\multicolumn{1}{|c|}{\textcolor{white}{\textbf{Method}}}
&
\multicolumn{1}{c|}{
\shortstack{
\textcolor{white}{\textbf{Stable Runs}}
}}
&
\multicolumn{1}{c|}{
\shortstack{
\textcolor{white}{\textbf{Stabilization Rate}}
}}
&
\multicolumn{1}{c|}{
\shortstack{
\textcolor{white}{\textbf{Median Stabilization Episode}}
}}
\\
\hline
Tabular \ac{SARSA}
& $29/30$
& $96.67\%$
& $850$ \\
\hline
Tabular $Q$-learning
& $28/30$
& $93.33\%$
& $550$ \\
\hline
Linear \ac{SARSA}
& $27/30$
& $90.00\%$
& $550$ \\
\hline
Linear $Q$-learning
& $26/30$
& $86.67\%$
& $550$ \\
\hline
\end{tabular}
\end{table}

\subsection{Main Benchmark Comparison at the $16\times8$ Setting} \label{Subsec:Main16x8Comparison}
\begin{figure*}[t]
\begin{tcolorbox}
\centering
\includegraphics[width=0.98\textwidth]{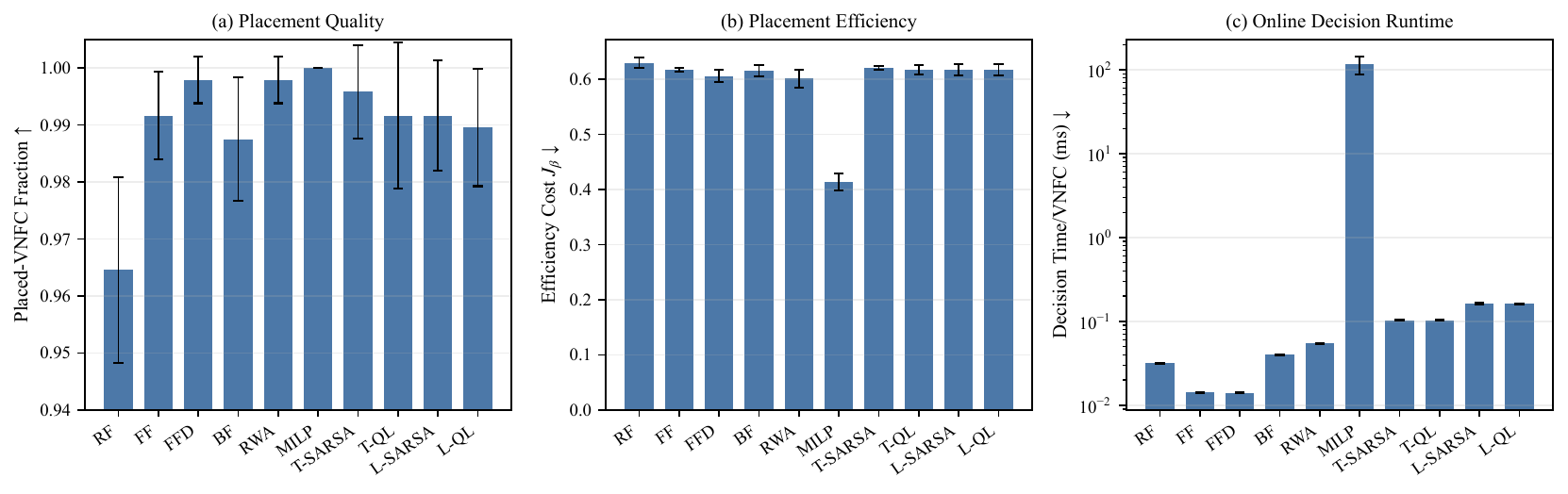}
\caption{Comparative analysis of the $16\times8$ configuration:
(a) Placement quality per placed-\ac{VNFC} fraction, (b) Efficiency cost $J_\beta$, and
(c) online decision time per \ac{VNFC} on a logarithmic scale. The error bars are normal-approximation $95\%$ confidence intervals over $30$ independent instances.}
\label{Fig:Main16x8Comparison}
\end{tcolorbox}
\vspace{-5.8mm}
\end{figure*}

We first compare all ten solution approaches on the reference $16\times8$ \ac{VNFC}-to-\ac{VM} configuration. Figure~\ref{Fig:Main16x8Comparison} summarizes placed-\ac{VNFC} fraction, efficiency cost $J_\beta$, and online decision time per \ac{VNFC}. Table~\ref{Tab:Main16x8Comparison} reports the corresponding numerical values with $95\%$ confidence intervals over the $30$ independent test instances. The results show that all four learning methods achieve high placement success, but they do not outperform the strongest conventional baselines on this static benchmark. Among the learning approaches, tabular \ac{SARSA} achieves the highest placed-\ac{VNFC} fraction, $0.996\pm0.008$. This is close to first-fit decreasing and resource-wastage-aware placement, both of which achieve $0.998\pm0.004$, while the \ac{MILP} baseline places all \acp{VNFC} in every evaluated instance.

The efficiency results reveal a larger separation. The \ac{MILP} baseline achieves the lowest mean efficiency cost, $J_\beta=0.414\pm0.015$, and the lowest active-\ac{VM} residual wastage, $0.048\pm0.011$. Among the non-optimization methods, the resource-wastage-aware heuristic obtains the lowest mean $J_\beta$, $0.601\pm0.016$, followed closely by first-fit decreasing at $0.606\pm0.011$. The four learning methods lie in the comparatively narrow range $J_\beta\approx0.617$--$0.621$. Accordingly, their high placement fractions do not translate into better resource consolidation than the strongest heuristics.

The runtime results exhibit the opposite trade-off. First-fit and first-fit decreasing require approximately $0.014$ ms per \ac{VNFC}, while the four learned policies require approximately $0.10$--$0.16$ ms per \ac{VNFC}. These learned-policy decision times remain small in absolute terms, but they are higher than those of the simplest heuristics. The \ac{MILP} baseline is substantially more expensive, requiring approximately $115.9$ ms per \ac{VNFC} on average under the solver configuration used here.

Overall, reference experiment does not indicate universal superiority of \ac{RL}. Rather, it shows a clear quality-latency trade-off: learning methods provide high placement success and low online decision latency after training, strong heuristics match or slightly exceed their placement quality at lower decision cost, and the \ac{MILP} baseline provides strongest static resource efficiency at substantially higher computational expense.

\subsection{Placement Quality and Computational Scaling} \label{Subsec:ScalabilityAnalysis}
We next examine how placement quality and computational cost evolve as the controlled problem size increases from $8\times4$ to $64\times32$ \ac{VNFC}-to-\ac{VM} configurations. The target aggregate resource load is kept fixed at $0.75$, so the comparison reflects increasing problem dimension rather than a progressively more over-provisioned workload.

\begin{table*}[t]
\caption{Comparative analysis of $16\times8$ setup. Values are means $\pm$ normal-approximation $95\%$ confidence half-widths over $30$ independent instances.}
\label{Tab:Main16x8Comparison}
\centering
\scriptsize
\renewcommand{\arraystretch}{1.40}
\setlength{\tabcolsep}{5pt}
\begin{tabular}{|l|c|c|c|c|}
\hline
\rowcolor{tableheader}
\multicolumn{1}{|c|}{\textcolor{white}{\textbf{Method}}}
&
\textcolor{white}{\textbf{Placed-VNFC Fraction}}
&
\textcolor{white}{$\boldsymbol{J_\beta}$}
&
\textcolor{white}{\textbf{Active-VM Wastage}}
&
\textcolor{white}{\textbf{Decision Time (ms/VNFC)}}
\\
\hline
Random feasible
& $0.965 \pm 0.016$
& $0.630 \pm 0.010$
& $0.265 \pm 0.014$
& $0.0316 \pm 0.0004$
\\
\hline
First-fit
& $0.992 \pm 0.008$
& $0.617 \pm 0.003$
& $0.235 \pm 0.006$
& $0.0142 \pm 0.0002$
\\
\hline
First-fit decreasing
& $0.998 \pm 0.004$
& $0.606 \pm 0.011$
& $0.221 \pm 0.011$
& $0.0141 \pm 0.0001$
\\
\hline
Best fit
& $0.988 \pm 0.011$
& $0.615 \pm 0.010$
& $0.235 \pm 0.012$
& $0.0398 \pm 0.0004$
\\
\hline
Resource-wastage-aware
& $0.998 \pm 0.004$
& $0.601 \pm 0.016$
& $0.221 \pm 0.015$
& $0.0547 \pm 0.0009$
\\
\hline
\ac{MILP}
& $1.000 \pm 0.000$
& $0.414 \pm 0.015$
& $0.048 \pm 0.011$
& $115.9 \pm 28.8$
\\
\hline
Tabular \ac{SARSA}
& $0.996 \pm 0.008$
& $0.621 \pm 0.004$
& $0.242 \pm 0.007$
& $0.1029 \pm 0.0011$
\\
\hline
Tabular $Q$-learning
& $0.992 \pm 0.013$
& $0.618 \pm 0.008$
& $0.240 \pm 0.010$
& $0.1032 \pm 0.0017$
\\
\hline
Linear \ac{SARSA}
& $0.992 \pm 0.010$
& $0.617 \pm 0.010$
& $0.244 \pm 0.010$
& $0.1646 \pm 0.0038$
\\
\hline
Linear $Q$-learning
& $0.990 \pm 0.010$
& $0.618 \pm 0.010$
& $0.245 \pm 0.010$
& $0.1613 \pm 0.0021$
\\
\hline
\end{tabular}
\vspace{-2.0mm}
\end{table*}

\begin{figure*}[t]
\begin{tcolorbox}
\centering
\includegraphics[width=1.0\textwidth]{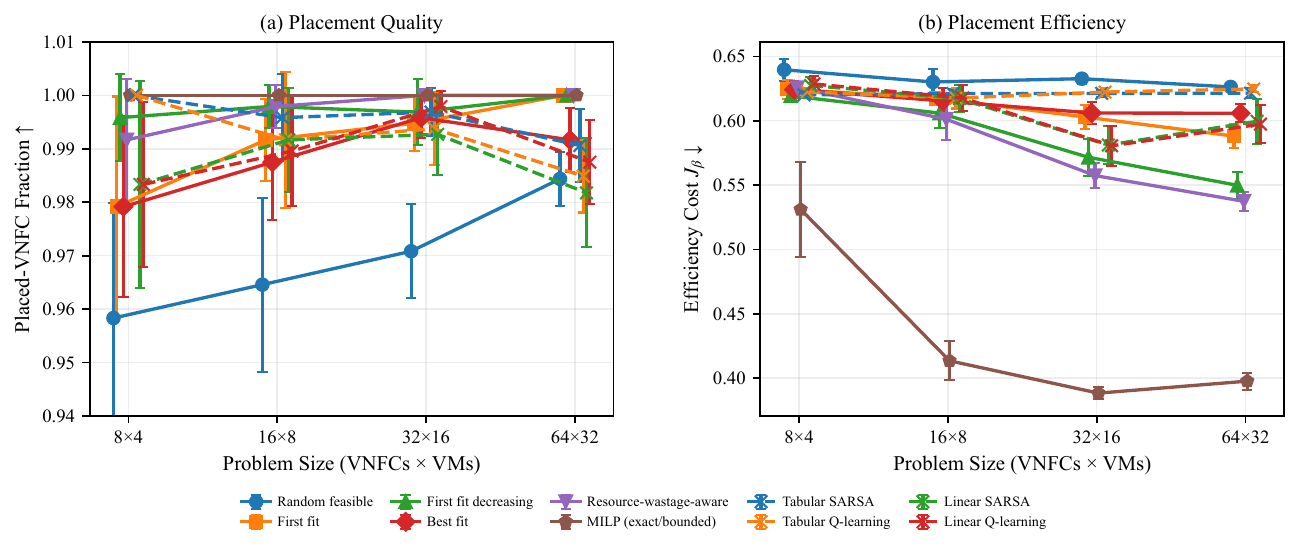}
\caption{(a) Placement quality per placed-\ac{VNFC} fraction, and (b) Efficiency cost $J_\beta$. Results are reported over $30$ independent test instances per size. The lower the efficiency cost, the better. The error bars are normal-approximation $95\%$ confidence intervals.}
\label{Fig:ScalabilityQuality}
\end{tcolorbox}
\vspace{-5.8mm}
\end{figure*}

Figure~\ref{Fig:ScalabilityQuality} shows the placed-\ac{VNFC} fraction and efficiency cost $J_\beta$ across the four controlled problem sizes. The placed-\ac{VNFC} fraction remains high for all methods over the tested range, although differences become more visible as the problem size increases. At the largest $64\times32$ configuration, first-fit decreasing, resource-wastage-aware placement, and the \ac{MILP} baseline achieve mean placed-\ac{VNFC} fractions of $\sim 1.000$. The learning methods remain close to full placement but do not uniformly match these strongest baselines.

The efficiency metric produces a clearer separation among the methods. At $64\times32$, the \ac{MILP} baseline obtains an efficiency cost of approximately $0.398$, while first-fit decreasing and the resource-wastage-aware heuristic achieve approximately $0.550$ and $0.537$, respectively. The learning methods remain in the approximate range $0.597$--$0.625$. Thus, increasing problem size does not reveal an efficiency advantage for \ac{RL} methods over strongest heuristic or optimization baselines.

\begin{figure*}[t]
\begin{tcolorbox}
\centering
\includegraphics[width=1.0\textwidth]{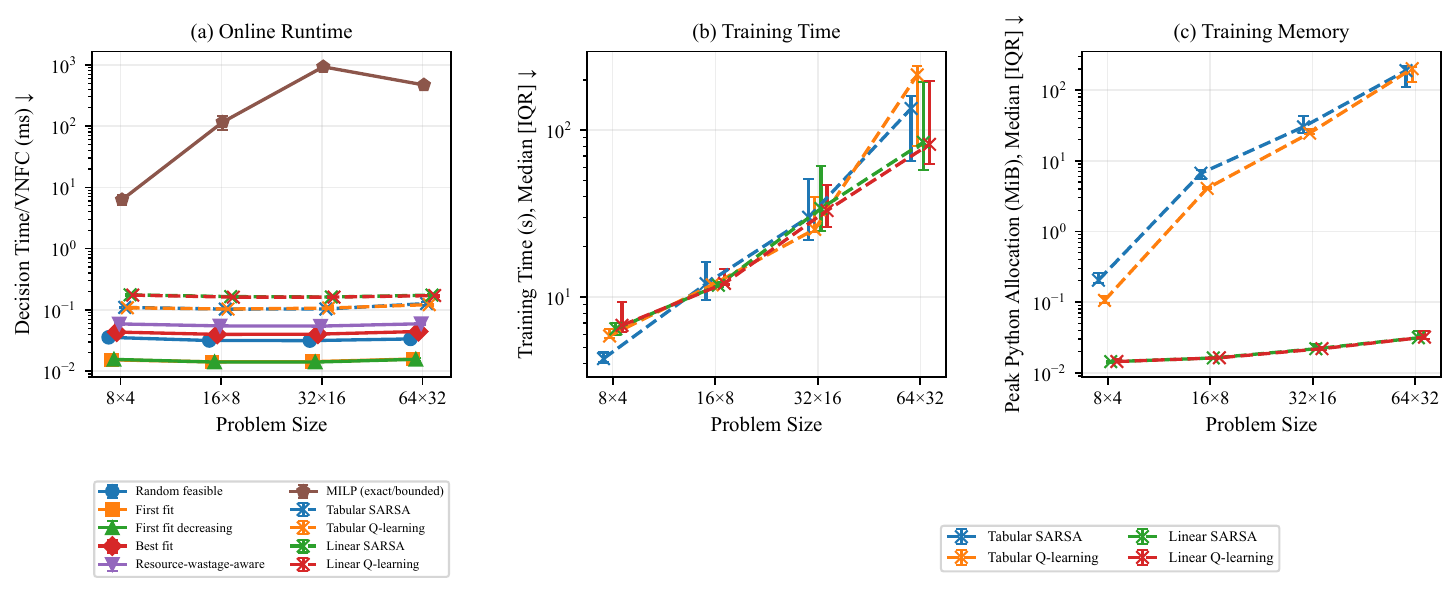}
\caption{Computational scaling with problem size:
(a) Online runtime per \ac{VNFC},
(b) \ac{RL} training time, and
(c) peak Python-managed allocation during \ac{RL} training.
Training time and peak allocation are
summarized using the median and interquartile range, while the decision time
is summarized across independent test instances. Decision times are the respective means with normal-approximation $95\%$ confidence intervals over $30$ instances per size, and the training time and peak allocation are medians with interquartile ranges.}
\label{Fig:ComputationalScaling}
\end{tcolorbox}
\vspace{-5.8mm}
\end{figure*}

Figure~\ref{Fig:ComputationalScaling} presents the corresponding computational scaling. Online decision time is reported for all methods, whereas training time and peak Python allocation apply only to the four learning methods. The online decision time of the four learned policies remains below approximately $0.18$ ms per \ac{VNFC} across all tested sizes. This should not be interpreted as dimension-independent computational complexity, since the largest action set considered contains only $32$ candidate \acp{VM} and each policy still evaluates candidate actions during decision-making. The $\sim 469$~ms/\ac{VNFC} value reported for \ac{MILP} at $64\times32$ is the $30$~seconds incumbent divided by $64$ requests. It should not be interpreted as improved scaling compared against $32\times16$ scenario.

The offline training cost increases substantially. For tabular \ac{SARSA}, the median training time rises from $4.28$ seconds at $8\times4$ to $134.74$ seconds at $64\times32$. For tabular $Q$-learning it increases from $5.85$ seconds to $214.08$ seconds. The linear methods increase from $6.5$--$6.8$ seconds at the smallest setting to approximately $82$--$85$ seconds at the largest setting.

The strongest representation-scaling difference is observed in memory. At $64\times32$, tabular \ac{SARSA} and tabular $Q$-learning require median allocations of approximately $189.6$ MiB and $201.8$ MiB, respectively, whereas both linear methods remain near $0.03$ MiB. These values correspond only to allocations observed through \texttt{tracemalloc} and should not be interpreted as total process memory.

\begin{figure}[t]
\begin{tcolorbox}
\centering
\includegraphics[width=\columnwidth]{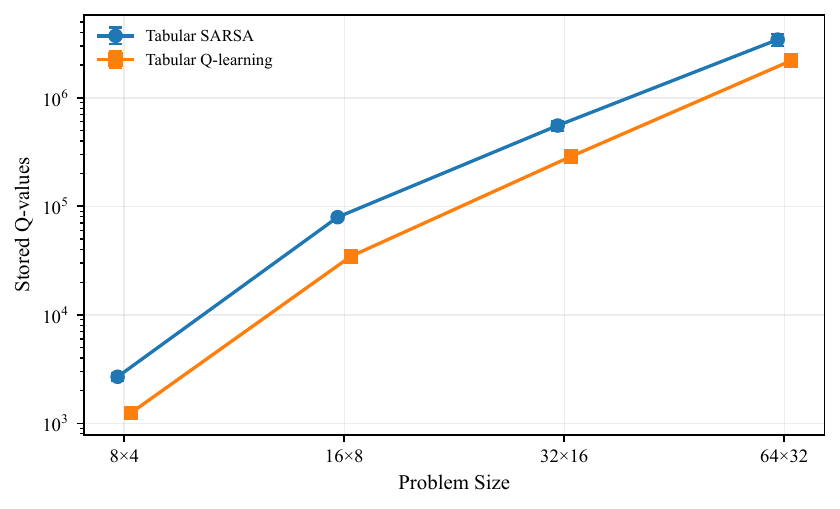}
\caption{Growth of the stored tabular state--action representation with
problem size. Only tabular \ac{SARSA} and tabular $Q$-learning methods are shown because the linear counterparts retain a fixed $15$-parameter representation.}
\label{Fig:TabularRepresentationGrowth}
\end{tcolorbox}
\vspace{-2.8mm}
\end{figure}

The growth of tabular representations is shown separately in Figure~\ref. At the $8\times4$ configuration, tabular \ac stores approximately $2.69\times10^3$ state-action values and tabular $Q$-learning approximately $1.25\times10^3$. At $64\times32$, these quantities increase to approximately $3.44$ million and $2.19$ million, respectively. In contrast, the linear methods retain the same 15 learned parameters at every problem size. This difference illustrates the principal scalability advantage of linear representation methods. Its learned representation size does not grow with the number of encountered discrete state-action pairs.

\subsection{Failure Conditions and Distribution Shift} \label{Subsec:FailureDistributionShift}
We next examine the behavior of the considered methods under controlled capacity stress and distribution shifts. The goal of this experiment is not to claim broad out-of-distribution generalization, but to identify the conditions under which the placement methods begin to lose completion performance.

The evaluated scenarios include the in-distribution reference case, a resource-abundant setting, a resource-scarce setting, resource fragmentation, a globally infeasible setting, heavier unseen \ac{VNFC} demands, smaller unseen \ac{VM} capacities, \ac{O-CU}-heavy and \ac{O-DU}-heavy demand profiles, positive and negative storage-compute correlation, and an unseen problem size. The reference, abundant, scarce, and globally infeasible $16\times8$ scenarios use target aggregate loads $0.75$, $0.50$, $0.95$, and $1.20$, respectively. Heavier demands multiply both the request resources by a factor of $1.25$, the smaller capacities multiply both the \ac{VM} resources by a factor of $0.80$, the \ac{O-CU}-heavy multiplies the \ac{O-CU} compute demand by a factor of $1.50$; and the \ac{O-DU}-heavy multiplies the \ac{O-DU} storage demand by a factor of $1.50$. Positive correlation pairs the storage and compute demands in the same sorted order, whereas negative correlation reverses the compute order. The fragmentation scenario uses eight $(10,10)$ \acp{VM} and a fixed globally feasible but order-sensitive set of $16$ requests. The unseen-size case is $32\times16$ at the load of $0.75$.

\begin{figure*}[t]
\begin{tcolorbox}
\centering
\includegraphics[width=0.98\textwidth]{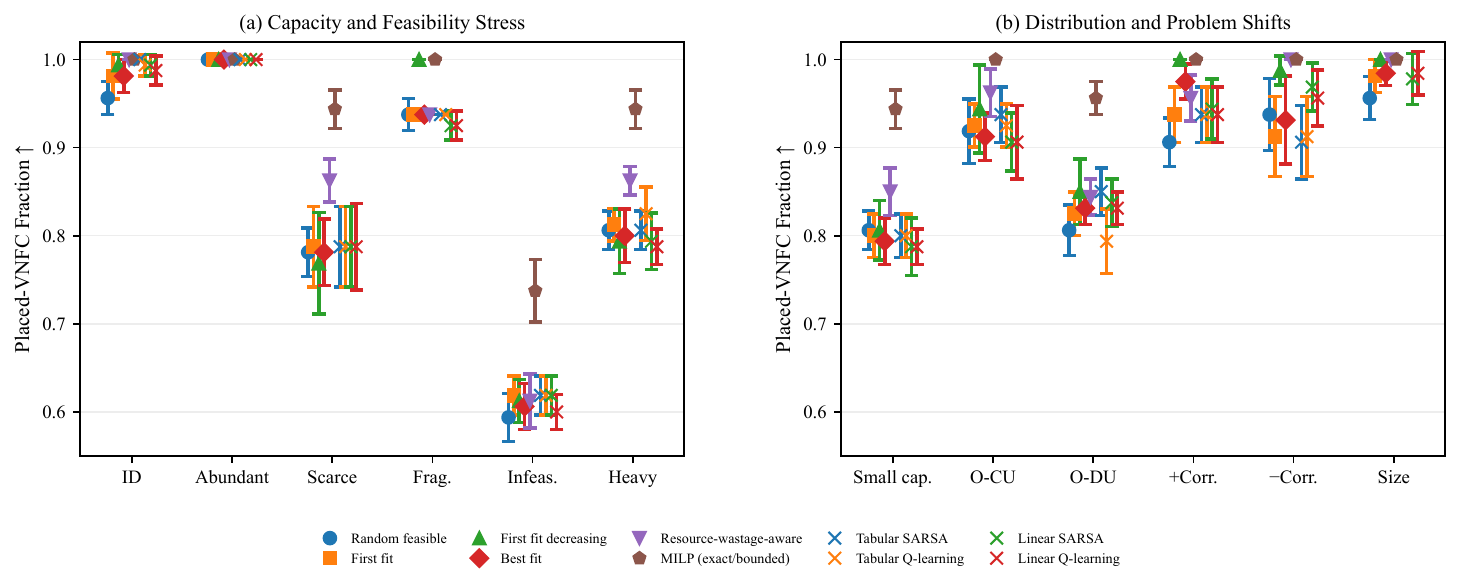}
\caption{Failure and distribution-shift analysis.
Panel~(a) considers capacity and feasibility stress, including the
in-distribution reference case, abundant resources, scarce resources,
fragmentation, global infeasibility, and heavier unseen demands.
Panel~(b) considers unseen smaller \ac{VM} capacities, \ac{O-CU}-heavy and
\ac{O-DU}-heavy demand profiles, positive and negative storage-compute
correlation, and an unseen problem size. Results are averaged over the
dedicated failure-analysis seeds. Error bars are normal-approximation $95\%$ confidence intervals over ten independent seeds.}
\label{Fig:FailureDistributionShift}
\end{tcolorbox}
\vspace{-4.8mm}
\end{figure*}

For the distribution-shift study in Figure \ref{Fig:FailureDistributionShift}, Plot (b), the learned models are trained on the corresponding $16\times8$ scenario and evaluated without retraining. For the unseen action dimension, only the linear models are evaluated. Tabular results are not applicable for unseen size because their stored action-value vectors have length $m$. Figure~\ref{Fig:FailureDistributionShift} reports the placed-\ac{VNFC} fraction under these scenarios. The results show that performance degradation is primarily associated with reduced feasibility margins rather than with every form of distribution change. In the resource-abundant case, all methods remain close to complete placement, whereas the resource-scarce, fragmentation, heavier-demand, and smaller-capacity scenarios produce visible reductions in placed-\ac{VNFC} fraction.

The globally infeasible scenario provides an important sanity check. Since the total resource configuration cannot support complete placement, no method can be expected to achieve a placed-\ac{VNFC} fraction of one. The reduced values in this setting therefore reflect problem infeasibility rather than a failure of a particular learning algorithm.

The \ac{O-CU}-heavy, \ac{O-DU}-heavy, and resource-correlation scenarios produce moderate changes in performance. This indicates that changing the composition and correlation structure of the request sequence does not necessarily cause catastrophic degradation when sufficient residual capacity remains.

The unseen problem-size case should be interpreted carefully. Linear methods can be evaluated directly because their learned representation remains fixed at $15$ parameters. In contrast, tabular methods rely on size-dependent discretized state keys and are not treated as having automatic generalization to unseen dimensions. This difference motivates reporting representational scalability separately from placement quality.

In summary, these results reinforce that the principal limitation of the sequential placement problem is feasibility under constrained residual resources. The experiments do not establish general out-of-distribution robustness, but they do show how each method behaves under specific controlled departures from the training distribution.

\subsection{Reward Ablation} \label{Subsec:RewardAblation}
The reward in Equation~\eqref{eq:revised-reward} separates successful placement from the terminal efficiency bonus. To assess whether the selected efficiency term materially improves the learned policy, we compare the final reward formulation against a feasibility-only variant using ten independent seeds.

\begin{figure*}[t]
\begin{tcolorbox}
\centering
\includegraphics[width=0.83\textwidth]{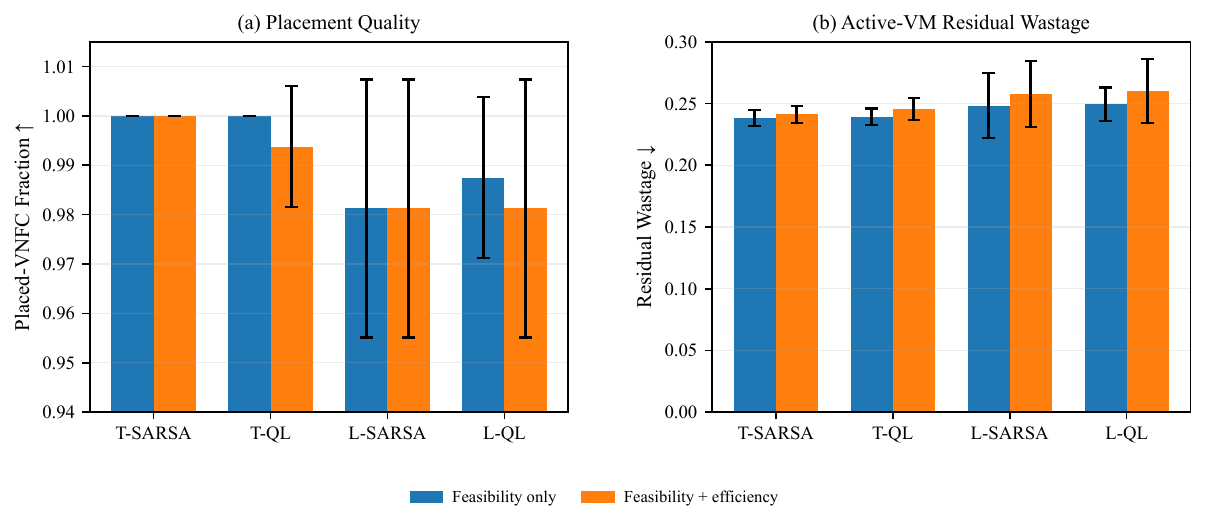}
\caption{Ablation of rewards on $16\times8$ configuration. Error bars are normal-approximation $95\%$ confidence intervals over ten independent seeds.}
\label{Fig:RewardAblation}
\end{tcolorbox}
\vspace{-5.8mm}
\end{figure*}

\begin{table*}[t]
\color{black}
\caption{Ablation of rewards on $16\times8$ configuration. Values are means $\pm$ normal-approximation 95\% confidence half-widths over ten independent seeds.}
\label{Tab:RewardAblation}
\centering
\scriptsize
\renewcommand{\arraystretch}{1.40}
\setlength{\tabcolsep}{5pt}
\begin{tabular}{|l|l|c|c|c|}
\hline
\rowcolor{tableheader}
\multicolumn{1}{|c|}{\textcolor{white}{\textbf{Method}}}
&
\multicolumn{1}{c|}{\textcolor{white}{\textbf{Reward}}}
&
\multicolumn{1}{c|}{\textcolor{white}{\textbf{Placed-VNFC Fraction}}}
&
\multicolumn{1}{c|}{\textcolor{white}{\textbf{Active-VM Wastage}}}
&
\multicolumn{1}{c|}{\textcolor{white}{$\boldsymbol{J_\beta}$}}
\\
\hline
Tabular \ac{SARSA}
& Feasibility only
& $1.000\pm0.000$
& $0.238\pm0.006$
& $0.619\pm0.003$
\\
& Feasibility + efficiency
& $1.000\pm0.000$
& $0.241\pm0.007$
& $0.621\pm0.003$
\\
\hline
Tabular $Q$-learning
& Feasibility only
& $1.000\pm0.000$
& $0.239\pm0.007$
& $0.620\pm0.003$
\\
& Feasibility + efficiency
& $0.994\pm0.012$
& $0.245\pm0.009$
& $0.623\pm0.004$
\\
\hline
Linear \ac{SARSA}
& Feasibility only
& $0.981\pm0.026$
& $0.248\pm0.027$
& $0.617\pm0.025$
\\
& Feasibility + efficiency
& $0.981\pm0.026$
& $0.258\pm0.027$
& $0.629\pm0.013$
\\
\hline
Linear $Q$-learning
& Feasibility only
& $0.988\pm0.016$
& $0.250\pm0.014$
& $0.625\pm0.007$
\\
& Feasibility + efficiency
& $0.981\pm0.026$
& $0.260\pm0.026$
& $0.630\pm0.013$
\\
\hline
\end{tabular}
\vspace{-4.0mm}
\end{table*}

Figure~\ref{Fig:RewardAblation} and Table~\ref{Tab:RewardAblation} summarize the results for the $16\times8$ configuration. The ablation does not show a systematic benefit from the tested terminal efficiency bonus with $\eta=0.25$. For tabular \ac{SARSA}, the placed-VNFC fraction remains unchanged at $1.000$, while $J_\beta$ increases slightly from $0.619$ to $0.621$. For tabular $Q$-learning, the placed-VNFC fraction decreases from $1.000$ to $0.994$, while $J_\beta$ increases from $0.620$ to $0.623$.

The same tendency is visible for the linear methods. Linear \ac{SARSA} retains the same mean placed-VNFC fraction of $0.981$, but its mean $J_\beta$ increases from $0.617$ to $0.629$. Linear $Q$-learning decreases from $0.988$ to $0.981$ in placed-VNFC fraction, while $J_\beta$ increases from $0.625$ to $0.630$.

Relative to the feasibility-only variant, the mean changes in $J_\beta$ are therefore approximately $+0.0015$, $+0.0031$, $+0.0120$, and $+0.0055$ for tabular \ac{SARSA}, tabular $Q$-learning, linear \ac{SARSA}, and linear $Q$-learning, respectively. Since lower $J_\beta$ is preferable, these changes provide no evidence that the selected terminal bonus improves placement efficiency.

\subsection{Limitations and Scope of the Evaluation} \label{Subsec:EvaluationLimitations}
The results should be interpreted within the scope of experimental design considered in this work. First, the evaluation addresses a static \ac{VNFC}-to-\ac{VM} placement problem in which finite sequence of \ac{VNFC} requests constitutes an episode. Dynamic arrivals, departures, migration, and time-varying \ac{VM} availability are not modeled. Consequently, the results do not establish performance under evolving \ac{O-Cloud} workloads.

Second, the controlled scalability experiments consider problem sizes up to $64\times32$ \ac{VNFC}-to-\ac{VM} configurations and a target aggregate load of $0.75$. Although these experiments expose differences in representation growth, training cost, and placement quality, they do not establish asymptotic scalability for substantially larger deployments.

Third, the distribution-shift experiments represent a collection of controlled stress cases rather than a comprehensive out-of-distribution benchmark. Performance under the considered shifts, therefore, should not be interpreted as evidence of universal generalization. This limitation is particularly important for the tabular methods, whose discretized state representation is intrinsically tied to the encountered state structure.

Fourth, the linear function approximation methods employ a manually designed $15$-dimensional feature representation. Their fixed representation size provides a substantial memory advantage over the tabular methods, but their performance depends on the information captured by these features.

Fifth, the main controlled benchmark is instance-specific. Upon selection of a method-level hyperparameter, a fresh model is trained and greedily evaluated on the same generated placement instance. Consequently, it measures learned optimization of a fixed batch and does not generalize to an independently generated instance.

Sixth, the simulated environment omits fronthaul/midhaul latency, transport connectivity, O-CU/O-DU communication dependencies, and other relevant \acp{KPI}. It models static compute and storage capacities only. As a result, the results do not establish real-world resilience or production deployment readiness.

Finally, the \ac{MILP} baseline is subject to a $30$ seconds solver limit. Results for instances in which the solver establishes optimality can be interpreted as solver-certified solutions, whereas results obtained at the time limit represent the best incumbent available within that budget. Solver status and relative \ac{MILP} gap are therefore retained.

Taken together, the experiments are intended to provide a controlled comparison of sequential \ac{VNFC}-placement strategies rather than to establish that one class of methods is universally preferable. The results characterize the trade-offs among placement feasibility, resource efficiency, online decision latency, offline training cost, and representation growth under the specified \ac{O-RAN}-oriented placement setting.

\section{Conclusion and Future Research Directions} \label{Sec:Conclusion}
In this paper, we formulated \ac{O-CU} and \ac{O-DU} \ac{VNFC}-to-\ac{VM} mapping problem as a capacity-constrained sequential decision-making problem. We evaluated tabular and linear variants of \ac{SARSA} and the tabular and linear variants of $Q$-learning against five conventional baselines and an \ac{MILP}. On $(h, m) = (16,8)$ setting, the highest mean placed-\ac{VNFC} fraction among the discussed \ac{RL} methods is $0.996\pm0.008$, compared against $0.998\pm0.004$ for resource-wastage-aware heuristic and first-fit decreasing and $1.000$ for \ac{MILP}. The \ac{RL} agents take decisions in approximately $0.103$-$0.165$ ms per \ac{VNFC}. Simultaneously, their efficiency costs are higher than those of the baselines.  Therefore, the temporal-difference learning methods provide fast online decisions compared with \ac{MILP}. Nevertheless, they do not outperform the strongest heuristics or the \ac{MILP} incumbents in placement quality or static resource efficiency. Their online latency must also be considered together with their nonzero offline training cost, while the heuristics require no training.

The scalability study exposes a key trade-off related to representation.  At $(64,32)$, tabular agents store as many as $2.19$--$3.44$ million $Q$ values, hence using approximately $190$-$202$ MiB of traced Python allocation. On the other hand, the $15$-parameter linear learning methods use approximately $0.03$ MiB. However, its performance quality on unseen scenarios and instances remains below compared to \ac{MILP} and strong heuristics. The ablation study of rewards finds no efficiency improvement from the tested terminal shaping weight. Moreover, the stress tests identify fragmentation, scarcity, infeasibility, and distribution shifts as clear limitations. These results support a quality-latency-memory trade-off and do not establish a universal superiority of any method.

In future, we aim to extend the current research from both a methodological perspective (i.e., by exploring more advanced \ac{RL} algorithms) and an applied perspective (i.e., applying the proposed approaches to practical use cases). First, we plan to implement deep $Q$-learning and double $Q$-learning algorithms with an aim to enhance the mapping of \ac{O-CU} and \ac{O-DU}, leveraging their ability to handle high-dimensional state spaces and improve convergence. Second, we intend to deploy these approaches in practical use cases, such as optimizing resource management, latency, energy consumption, and other \acp{KPI} in real-world scenarios. These efforts will further bridge the gap between theoretical frameworks and practical applications, enabling more efficient and adaptive solutions for \ac{O-RAN}.

\appendices

\newcolumntype{P}[1]{>{\centering\arraybackslash}m{#1}}
\newcolumntype{L}[1]{>{\raggedright\arraybackslash}m{#1}}

\begin{table*}[ht]
\centering
\caption{Summary of recent studies on VNF mapping, comparing objectives, methodologies, and KPIs in cellular and cloud/data center networks}
\renewcommand{\arraystretch}{1.4}

\resizebox{\textwidth}{!}{%
\begin{tabular}{|P{1cm}|L{3cm}|P{1.5cm}|L{7.2cm}|L{6.6cm}|L{4.6cm}|}
\hline \hline
\multicolumn{1}{|c|}{\cellcolor[HTML]{343434}\bfseries \textcolor{white}{Category}} &
\multicolumn{1}{c|}{\cellcolor[HTML]{343434}\bfseries \textcolor{white}{Reference}} &
\multicolumn{1}{c|}{\cellcolor[HTML]{343434}\bfseries \textcolor{white}{Year}} &
\multicolumn{1}{c|}{\cellcolor[HTML]{343434}\bfseries \textcolor{white}{Objectives}} &
\multicolumn{1}{c|}{\cellcolor[HTML]{343434}\bfseries \textcolor{white}{KPIs}} &
\multicolumn{1}{c|}{\cellcolor[HTML]{343434}\bfseries \textcolor{white}{Methodology}} \\
\hline

\multirow{13}{*}{%
  \rotatebox{90}{\parbox{7.6cm}{\centering
  Proposed approaches for \ac{VNF} mapping in cloud/data center networks}}}
  & Beloglazov \textit{et al}.\ \cite{6269025} & $2013$
  & Optimize \ac{VM} consolidation, minimize energy consumption
  & Energy consumption and \ac{VM} migrations
  & Markov chain model \\ \cline{2-6}

  & Xiao \textit{et al}.\ \cite{6311403} & $2013$
  & Prevent overload of \acp{PM} by effective \ac{VM} placement to support green computing
  & No. of overloaded servers, No. of actively used \acp{PM}, No. of \ac{VM} migrations, and energy savings
  & Fast up and slow down algorithm prediction \\ \cline{2-6}

  & Alameddine \textit{et al}.\ \cite{8052155} & $2017$
  & Optimize traffic routing and scheduling
  & Schedule length and execution time
  & Column generation with primal-dual decomposition \\ \cline{2-6}

  & Nejad \textit{et al}.\ \cite{8314726} & $2018$
  & Optimize \ac{VNF} placement and admission control
  & Acceptance rate and revenue from \ac{VNF} requests
  & Mixed integer linear programming \\ \cline{2-6}

  & Quang \textit{et al}.\ \cite{8873660} & $2019$
  & Optimize \ac{VNF} placement
  & Acceptance ratio, resource utilization, and \ac{QoS}
  & Enhanced exploration deep deterministic policy gradient  \\ \cline{2-6}

  & Pei \textit{et al}.\ \cite{8932445} & $2020$
  & Minimize cost and improve performance in \ac{VNF} placement
  & Request reject rate, throughput delay, and \ac{VNF} utilization
  & Double deep $Q$ network \\ \cline{2-6}

  & Pham \textit{et al}.\ \cite{7859379} & $2020$
  & Minimize operational and traffic costs for \ac{VNF} service chain placement
  & Operational energy consumption, network latency, bandwidth usage, and total cost
  & Markov approximation and matching theory \\ \cline{2-6}

  & Azizi \textit{et al}.\ \cite{9129768} & $2021$
  & Minimize power consumption and resource wastage
  & Active \acp{PM}, resource wastage, and power consumption
  & Greedy randomized algorithm \\ \cline{2-6}

  & Sallam \textit{et al}.\ \cite{9361423} & $2021$
  & Maximize processed network flows under realistic constraints
  & Processed flow volume, budget usage, and capacity utilization
  & Integer programming and submodular optimization algorithms \\ \cline{2-6}

  & Kuai \textit{et al}.\ \cite{9906071} & $2022$
  & Maximize fairness in \ac{VNF} mapping and scheduling across service function chains satisfying delay requirements
  & Minimum earliness and acceptance ratio
  & Proximal policy optimization and deep \ac{RL} \\ \cline{2-6}

  & Seyyedsalehi \textit{et al}.\ \cite{9870779} & $2022$
  & Optimize \ac{VM} placement in multi-data center clouds for big data tasks
  & Energy consumption, intra-data center traffic, \ac{VM} migration, and service level agreement violations
  & Aware genetic algorithm first fit (AGAFF) algorithm \\ \cline{2-6}

  & Bagaa \textit{et al}.\ \cite{9305277} & $2022$
  & Optimize security orchestration considering \ac{QoS} and heterogeneous resource constraints
  & End-to-end delay, bandwidth, resource utilization, and security levels
  & Linear programming and heuristic algorithms \\ \cline{2-6}

  & Wei \textit{et al}.\ \cite{10946227} & $2025$
  & Adaptively map service function chains to minimize cost and network congestion
  & Service delay, bandwidth consumption, network load balancing, and acceptance ratio
  & Deep \ac{RL} with long short-term memory \\ \hline

\multirow{6}{*}{%
  \rotatebox{90}{\parbox{4cm}{\centering
  Proposed approaches for \ac{VNF} mapping}}}
  & Mijumbi \textit{et al}.\ \cite{7849149} & $2017$
  & Predict \ac{VNF} resource needs using topology-aware learning approaches
  & Predict accuracy, call drop rate, latency, and resource efficiency
  & Graph neural networks \\ \cline{2-6}

  & Domenico \textit{et al}.\ \cite{9177288} & $2020$
  & Optimize \ac{VNF} deployment and resource allocation in hybrid cloud infrastructure
  & Resource usage, No.\ of supported slices, and latency complications
  & Integer linear programming \\ \cline{2-6}

  & Mei \textit{et al}.\ \cite{9261244} & $2020$
  & Optimize embedding of \ac{5G} slices with sharable \acp{VNF} to improve resource utilization
  & Slice acceptance ratio and physical resource utilization
  & Integer linear programming and heuristic algorithm \\ \cline{2-6}

  & Kist \textit{et al}.\ \cite{9302630} & $2022$
  & Enable RAN slicing with isolation, programmability, adaptability, and scalability
  & Slice isolation, performance, scalability and customization flexibility
  & Prototype combining baseband unit and remote radio head hypervisors \\ \cline{2-6}

  & Tran \textit{et al}.\ \cite{10623413} & $2024$
  & Optimize network throughput by maximizing the No. of service function chains embedded in the network
  & Acceptance ratio, network throughput ratio, and execution time
  & Deep-\ac{RL} (specifically deep $Q$-learning and advantage actor critic approaches) \\ \cline{2-6}

  & Hojeij \textit{et al}.\ \cite{10712647} & $2025$
  & \ac{O-CU} and \ac{O-DU} placement in \ac{O-RAN}
  & Maximizing user admittance, minimizing deployment costs, and increasing throughput
  & Recurrent neural network \\ \hline

Our paper
  & Habibi \textit{et al}. & $2026$
  & Mapping the \acp{VNFC} of \ac{O-CU} and \ac{O-DU} within an \ac{O-RAN} slice subnet onto \acp{VM} within \acp{O-Cloud} in \ac{O-RAN}
  & Minimize resource wastage while ensuring efficient and scalable \ac{VNFC} mapping
  & \ac{RL}-based methods \\ \hline
\end{tabular}}%

\label{tab:vnf_mapping}
\vspace{-4mm}
\end{table*}


\begin{appendices}
\section{Review of Existing State-of-the-Art Approaches to the Mapping Problem} \label{App:StateOfArt}
The \ac{VNF} mapping problem has been extensively investigated in both cloud/data center networks and wireless networks using a wide range of methodological approaches. This appendix surveys the most recent and relevant contributions in the literature, with particular emphasis on the key challenges addressed by existing solutions. A structured overview of selected works is presented in Table \ref{tab:vnf_mapping}, which compares the studies in terms of their objectives, employed methodologies, and the \acp{KPI} they optimize.

For clarity and systematic presentation, the reviewed literature is categorized into two groups: (i) approaches proposed for \ac{VNF} mapping in cellular networks and (ii) approaches developed for cloud/data center networks. The table consolidates the analyzed works by summarizing their objectives, methodologies, and optimized \acp{KPI}, thereby highlighting both the diversity of proposed solutions and the specific \ac{VNF} mapping challenges they address.


\section{Justification for Decomposing \ac{O-CU} and \ac{O-DU} into \acp{VNFC} for Enhanced \ac{O-RAN} Slicing} \label{App:VNFCJustification}
We propose that decomposing \ac{O-CU} and \ac{O-DU} into \acp{VNFC} offers a promising approach for achieving fine-grained resource allocation, enhancing \ac{O-RAN} slicing, and enabling more efficient mapping of \acp{VNFC} in \ac{O-CU} and \ac{O-DU} onto \acp{VM} within \ac{O-Cloud} sites. This appendix outlines the advantages of this proposal, illustrating how splitting \acp{VNF} into \acp{VNFC} can improve operational flexibility, scalability, and fault tolerance while simultaneously optimizing the performance of \ac{O-RAN} slices.

\textbf{Enhanced \ac{O-RAN} Slicing Flexibility:} By decomposing \ac{O-CU} and \ac{O-DU} into \acp{VNFC}, each \ac{VNFC} can be individually scaled and allocated to specific \ac{O-RAN} slices based on their functional requirements or predefined policies. For instance, certain \ac{PDCP} functionalities may introduce significant latency and therefore need to be disabled for \ac{O-RAN} slices providing \ac{URLLC} services. Similarly, some \ac{RLC} functionalities, such as segmentation, might only be enabled for \ac{eMBB} slices, while being disabled for \ac{URLLC} and \ac{mMTC} slices \cite{8931318}. An additional benefit of this approach is that individual \acp{VNFC} can be instantiated, migrated, or modified without disrupting the entire \ac{O-CU} or \ac{O-DU}, enabling a more agile response to slice-specific needs in \ac{O-RAN}.

\textbf{Finer Control over Workload Distribution and Hardware Utilization:} Splitting into \acp{VNFC} provides finer control over workload distribution across different computing nodes or \ac{O-Cloud} sites. For example, \acp{VNFC} requiring low latency can be deployed closer to the network edge, while those with less stringent latency demands may reside in centralized \ac{O-Cloud} sites. Moreover, \ac{VNFC}-level decomposition facilitates better utilization of heterogeneous hardware resources, such as \acp{GPU} for \ac{PHY}-layer \acp{VNFC} or general \acp{CPU} for higher-layer \acp{VNFC}. Taken together, these benefits significantly optimize \ac{VNF} mapping compared to traditional approaches.

\textbf{Improved Fault Tolerance and Resilience:} In this approach, if one \ac{VNFC} experiences issues, the impact is limited to the specific functionality it provides. Hence, this isolation enhances the resilience and reliability of an \ac{O-gNB}. Moreover, if a \ac{VNFC} fails, it can be restarted or replaced without affecting other components within the \ac{O-CU} or \ac{O-DU}.

\textbf{Facilitating Focused Optimization:} With distinct \acp{VNFC}, \ac{AI}/\ac{ML} models can optimize specific functionalities more effectively. For example, predicting congestion for the \ac{PDCP} \ac{VNFC} or improving scheduling in the \ac{MAC} \ac{VNFC}. Moreover, granular monitoring of \acp{VNFC} generates more detailed metrics, which can be collected and utilized by \ac{AI}/\ac{ML} models to enhance performance analysis of the \ac{O-CU} and \ac{O-DU}.

\textbf{Alignment with Microservice Architectures:} Our proposal is aligned with microservice-based architectural principles, which emphasize modularity and ease of management. We believe that a \ac{VNFC}-based architecture is more open to integrating functionalities, algorithms, and features from multiple vendors, adhering to the spirit of openness in \ac{O-RAN}. For example, within the \ac{O-CU}, an operator may select three \acp{VNFC} from different vendors, choosing each one based on its superior performance compared to competing \acp{VNFC} in the market. This approach promotes innovation and competition.

Despite these advantages, decomposing the \ac{O-CU} and \ac{O-DU} into \acp{VNFC} presents several challenges that must be addressed to fully realize its potential, as discussed below.

\textbf{Increased Overhead:} Managing multiple \acp{VNFC} may introduce additional overhead in signaling, orchestration, and management of a \ac{VNF}. Each \ac{VNFC} requires its own control and monitoring mechanisms, increasing intercomponent communication for tasks such as state synchronization, fault detection, and load balancing. To mitigate this overhead, efficient lifecycle management strategies are needed, such as selectively instantiating specific \acp{VNFC} only when required and leveraging lightweight virtualization technologies (e.g., containers) to reduce the operational footprint.

\textbf{Complexity:} Decomposing the \ac{O-CU} and \ac{O-DU} can introduce complexity at multiple levels. First, managing a larger number of \ac{VNFC} instances places additional demands on orchestration and automation frameworks (e.g., \ac{NFV-MANO}), which must dynamically schedule and maintain these microservices across distributed \ac{O-Cloud} sites. Second, cross-layer dependencies become more challenging, as individual \acp{VNFC} may require frequent state synchronization, resource sharing, or common security policies. Finally, this approach can complicate debugging, since identifying the root cause of performance issues or outages often involves tracing interactions across numerous microservices. Addressing these complexities necessitates robust management and orchestration solutions at the \ac{VNFC}, \ac{VNF}, and \ac{O-gNB} levels.


\section{Evaluating the Sequential Decision-Making Nature of Mapping O-CU and O-DU onto O-Clouds} \label{App:IsMappingaSDMP?}
The problem we address in this article, mapping a given \ac{VNFC} onto a \ac{VM}, is fundamentally a sequential decision-making optimization problem. This means that, rather than making all decisions simultaneously, the process unfolds iteratively, with one decision made at a given time. Each choice influences the subsequent decisions due to resource availability changes, dependencies between the \acp{VNF}, and the state of the underlying \acp{O-Cloud}. This approach makes the sequence of actions critical to achieving optimal outcomes.

At each step of the mapping process, the task is to assign a \ac{VNFC} to a suitable \ac{VM}. For example, \ac{VNFC} $f_1$ may be assigned to \ac{VM} $v_j$ based on current resource availability and placement constraints. The process then proceeds iteratively to the assignment of \ac{VNFC} $f_{i}$, for $i \in \{2, \dots, h\}$.

The process continues with each \ac{VNFC} -- from $f_2$ to $f_h$ -- being placed one at a time, with the system state updating after each decision. Naturally, earlier allocation decisions constrain the options available for future assignments. For example, assigning a large \ac{VNFC} to a \ac{VM} with limited capacity early in the process may hinder the efficient placement of subsequent components, particularly if the remaining \acp{VM} lack sufficient resources to host the remaining \acp{VNFC}. This direct dependency, where earlier decisions influence the feasibility and quality of future ones, is a defining characteristic of a sequential decision-making problem.

Therefore, the \ac{VNFC}-to-\ac{VM} placement problem can be formulated as a sequential decision-making problem. This allows us to apply \ac{RL} techniques, such as \ac{SARSA} and $Q$-learning -- specifically through \texttt{Algorithm 1}, \texttt{Algorithm 2}, \texttt{Algorithm 3}, and \texttt{Algorithm 4} - to learn a policy that selects actions (i.e., placements) in a manner that optimizes overall performance.
\end{appendices}


\section*{Acknowledgment}
This research was partially supported by the German Federal Ministry of Research, Technology and Space (BMFTR) through the Open6GHub+ project (Grant No.~16KIS2402K) and 6G Terafactory project (Grant No.~16KISK186). The authors gratefully acknowledge the anonymous reviewers for their insightful comments and constructive feedback, which significantly improved the quality of this article. The authors also thank the Area Editor and the Associate Editor for their efforts in coordinating the peer-review process.
\vspace{-2mm}

\section*{Acronym List}
\begin{acronym}
\acro{5G}{fifth-generation}
\acro{6G}{sixth-generation}

\acro{AI}{artificial intelligence}
\acro{AUC}{area under the curve}

\acro{CPU}{central processing unit}

\acro{eMBB}{enhanced mobile broadband}
\acro{ETSI}{European Telecommunications Standards Institute}

\acro{GPU}{graphics processing unit}

\acro{KPI}{key performance indicator}

\acro{MAC}{medium access control}
\acro{MILP}{mixed-integer linear programming}
\acro{ML}{machine learning}
\acro{mMTC}{massive machine-type communications}

\acro{NFV-MANO}{network function virtualization-management and orchestration}
\acro{NFVO}{network function virtualization orchestrator}
\acro{Non-RT RIC}{non-real-time RAN intelligent controller}
\acro{NP}{nondeterministic polynomial time}

\acro{O-Cloud}{open-cloud}
\acro{O-CU}{open centralized unit}
\acro{O-DU}{open distributed unit}
\acro{O-gNB}{open next generation node B}
\acro{O-RAN}{open radio access network}
\acro{O-RU}{open radio unit}

\acro{PDCP}{packet data convergence protocol}
\acro{PHY}{physical}
\acro{PM}{physical machine}

\acro{QoS}{quality of service}

\acro{RLC}{radio link control}
\acro{RL}{reinforcement learning}
\acro{RRC}{radio resource control}

\acro{SARSA}{state action reward state action}
\acro{SDAP}{service data adaptation protocol}
\acro{SMO}{service management and orchestration}


\acro{URLLC}{ultra-reliable low latency communications}

\acro{VIM}{virtualized infrastructure manager}
\acro{VM}{virtual machine}
\acro{VNF}{virtual network function}
\acro{VNFC}{virtual network function component}
\acro{VNFM}{virtual network function manager}
\end{acronym}


\bibliography{mypaper01.bib} 
\bibliographystyle{ieeetr}
\vspace{-1.5mm}


\section*{Author Biographies}
\vspace{-1.5mm}
\begin{IEEEbiography}[{\includegraphics[width=1in,height=1.25in, clip,keepaspectratio]{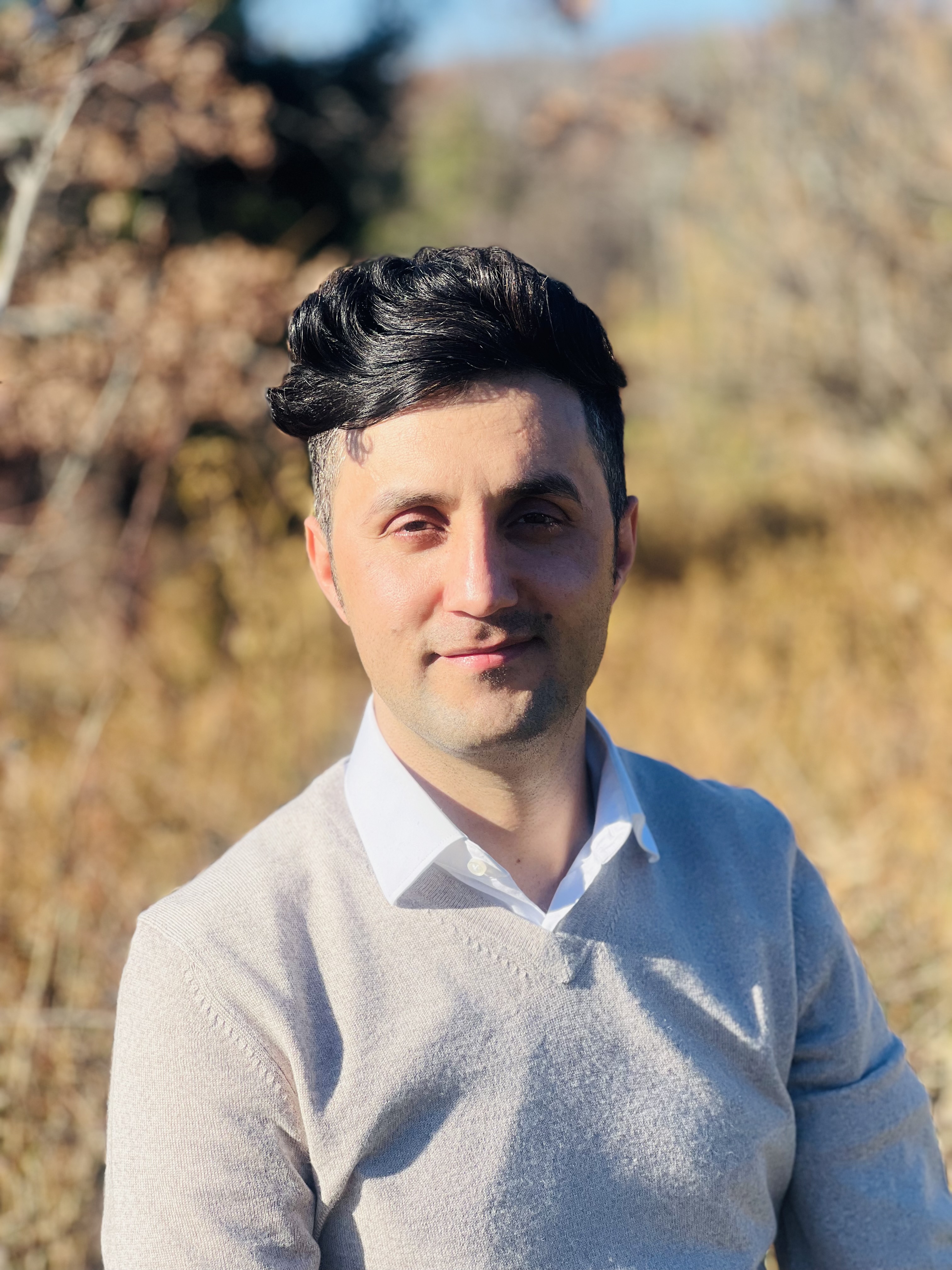}}]{\textsc{Mohammad Asif Habibi}} received the B.Sc. degree in Telecommunication Engineering from Kabul University, Afghanistan, in $2011$, the M.Sc. degree in Systems Engineering and Informatics from the Czech University of Life Sciences, Czech Republic, in $\mathrm{2016}$, and the Dr.-Ing. degree (Summa Cum Laude) from the University of Kaiserslautern (RPTU), Germany, in $2025$. From $2011$ to $2014$, he worked as a Radio Access Network Engineer at Huawei. From $2017$ to $2025$, he was a Researcher and Senior Researcher with the Division of Wireless Communications and Radio Navigation, University of Kaiserslautern (RPTU), Germany. Since $2026$, he has been a Senior Researcher at the German Research Center for Artificial Intelligence (DFKI). His research interests include network and service management, network architecture, machine learning, and radio access networks.
\end{IEEEbiography} 

\begin{IEEEbiography}[{\includegraphics[width=1.5in, height=1.25in,clip,keepaspectratio]{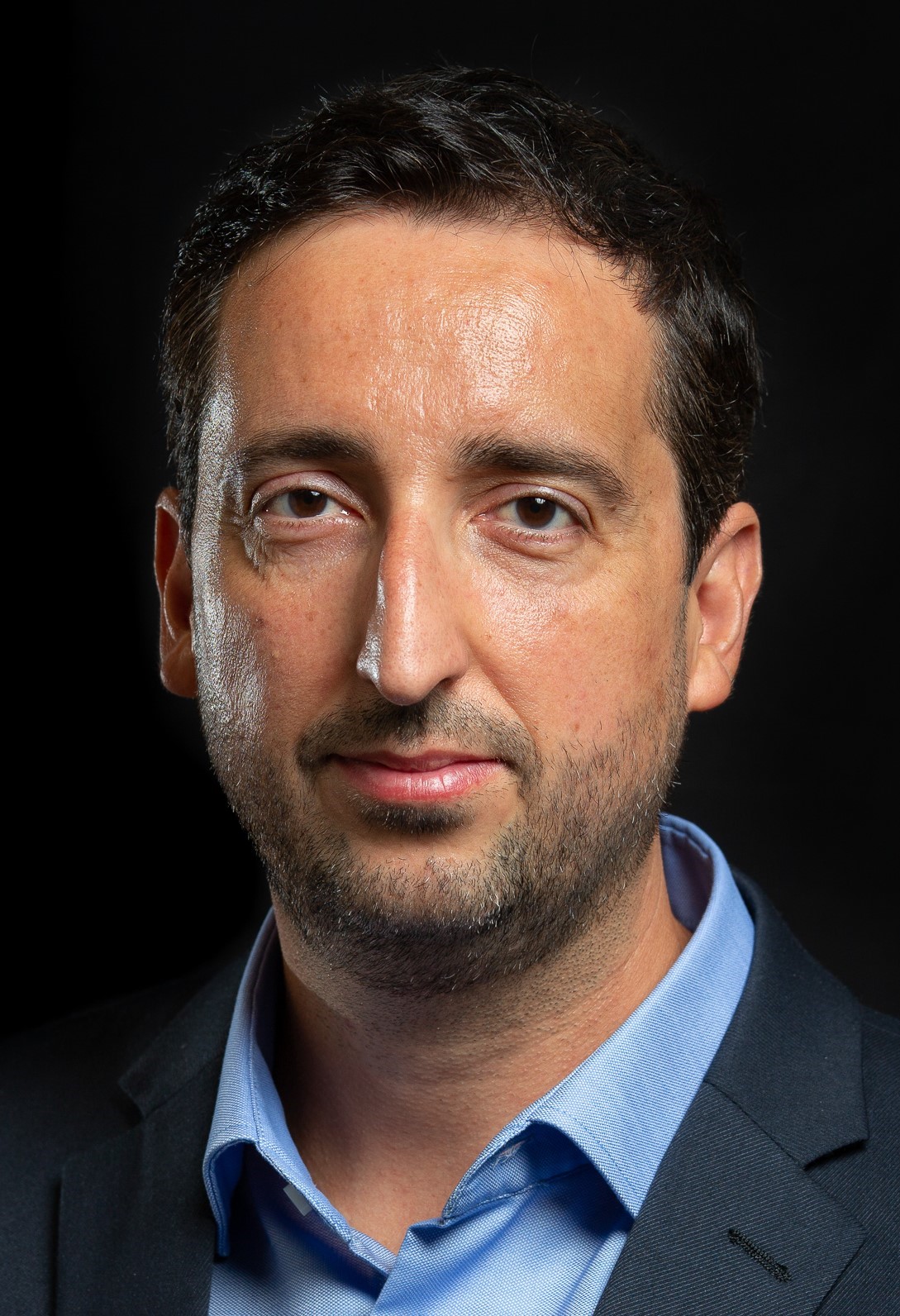}}]{\textsc{Xavier Costa-P\'erez}} (M'$\mathrm{06}$--SM'$\mathrm{18}$)~is an ICREA Research Professor and Scientific Director at the i2cat Research Center, concurrently leading Space and Terrestrial Systems (ASTRO) R\&D at NEC Laboratories Europe. A globally recognized leader in his field, Dr. Costa-Pérez has been ranked among the Top 2\% of World Scientists by Stanford for over 5 years and is featured on the Research.com Best World Computer Scientists list. His research spearheads the digital transformation of society through the convergence of Space and Mobile Networks together with Artificial Intelligence. Notably, his innovations have been integrated into commercial mobile phones, base stations, and network management systems, and have spurred the creation of multiple start-ups. He is a recognized contributor to multiple standard specifications who actively participated in bodies such as IEEE 802.11, 802.16, WiFi Alliance, and 3GPP. He has also served on organizing committees for prominent conferences (e.g., AAAI, ACM MOBICOM, IEEE INFOCOM) and as an Editor for leading journals like IEEE Transactions on Mobile Computing (TMC), IEEE Transactions on Communications (TCOM), as well as Elsevier Computer Communications (COMCOM). Dr. Costa-Pérez holds M.Sc. and Ph.D. degrees in Telecommunications from the Polytechnic University of Catalonia (UPC), receiving a national award for his doctoral thesis. He is the inventor of over 100 granted patents, some standard essential (SEP).
\end{IEEEbiography}

\begin{IEEEbiography}[{\includegraphics[width=1in, height=1.25in, clip, keepaspectratio]{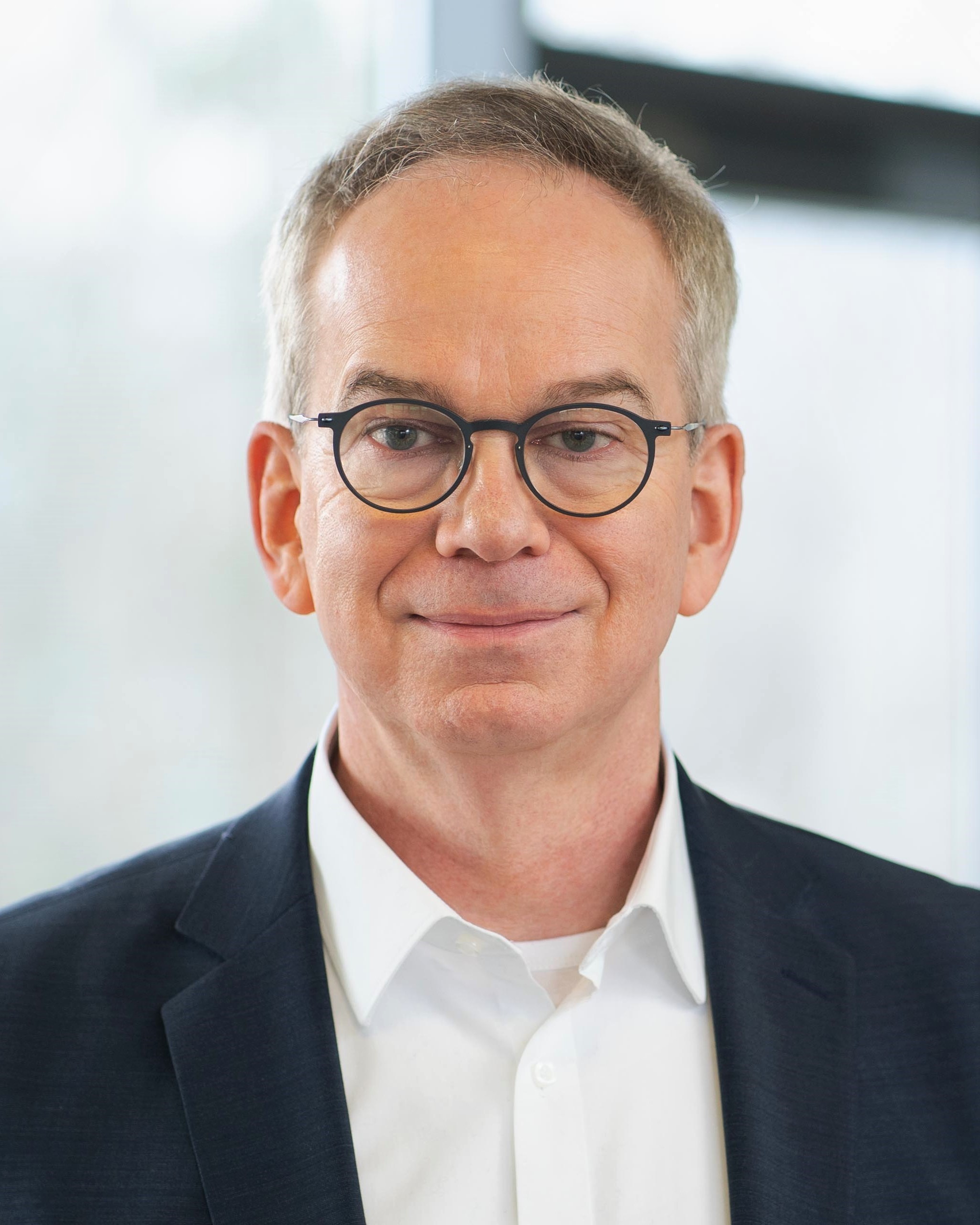}}]{\textsc{Hans D. Schotten}} (S'$\mathrm{93}$--M'$\mathrm{97}$) received the Diploma and Ph.D. degrees in electrical engineering from the Aachen University of Technology, Germany, in $\mathrm{1990}$ and $\mathrm{1997}$, respectively. Since August $\mathrm{2007}$, he has been a full professor and head of the Division of Wireless Communications and Radio Navigation at University of Kaiserslautern (RPTU). Since $\mathrm{2012}$, he has also been Scientific Director at the German Research Center for Artificial Intelligence, heading the Intelligent Networks department. He was a senior researcher, the project manager, and the head of the research groups at Aachen University of Technology, Ericsson Corporate Research, and Qualcomm Corporate R\&D. During his time at Qualcomm, he has also been the Director for Technical Standards and Coordinator of Qualcomm’s activities in European research programs.
\end{IEEEbiography}    

\end{document}